\documentclass[11pt]{article}
\usepackage{latexsym}
\usepackage{theorem}
\usepackage{graphicx}

\usepackage{amsmath,color}
\usepackage{amsfonts}
\usepackage[labelformat=simple]{subcaption}

\usepackage{natbib}
\usepackage{soul}
\usepackage{tikz}
\usetikzlibrary{arrows,shapes,positioning,arrows,fit,matrix}

\usepackage{geometry} 

\usepackage{pgfplots}
\pgfplotsset{compat=newest}

\tikzset
{
    treenode/.style = {circle, draw=black, align=center, minimum size=1cm},
}

\theorembodyfont{\rmfamily}
\newtheorem{theorem}{Theorem}

\newtheorem{lemma}[theorem]{Lemma}

\newtheorem{corollary}[theorem]{Corollary}

\theoremstyle{break}

\title{A General Framework for Approximating \\Min Sum Ordering Problems}

\author{Felix Happach\thanks{Department of Mathematics and TUM School of Management, Technische Universitat M\"{u}nchen, 80333 M\"{u}nchen, Germany, felix.happach@tum.de} \and Lisa Hellerstein\thanks{Department of Computer Science and Engineering, New York University Tandon School of Engineering, New York, New York 11201, lisa.hellerstein@nyu.edu} \and Thomas Lidbetter\thanks{Department of Management Science and Information Systems, Rutgers Business School, Newark, NJ 07102, tlidbetter@business.rutgers.edu}}

\providecommand{\keywords}[1]{\textbf{\textbf{Keywords:}} #1}

\begin{document}

\maketitle

\begin{abstract}
\noindent
We consider a large family of problems in which an ordering (or, more precisely, a chain of subsets) of a finite set must be chosen to minimize some weighted sum of costs.  This family includes variations of Min Sum Set Cover (MSSC), several scheduling and search problems, and problems in Boolean function evaluation.  We define a new problem, called the Min Sum Ordering Problem (MSOP) which generalizes all these problems using a cost and a weight function defined on subsets of a finite set. Assuming a polynomial time $\alpha$-approximation algorithm for the problem of finding a subset whose ratio of weight to cost is maximal, we show that under very minimal assumptions, there is a polynomial time $4 \alpha$-approximation algorithm for MSOP. This approximation result generalizes a proof technique used for several distinct problems in the literature. We apply this to obtain a number of new approximation results.
\end{abstract}


\newpage

\section{Introduction}
\label{sec:introduction}

Many optimization problems require finding a minimum cost (feasible) subset of the elements of a finite set, according to some cost function. For example, in the Set Cover Problem, the given finite set is a collection of subsets of a finite set of ``ground'' elements $X$, and the objective is to choose a sub-collection of minimum cardinality whose union is $X$. 
In the Minimum Spanning Tree Problem, for a given graph with edge lengths, the objective is to find a subtree of minimum total length that contains all the vertices of the graph. 

This work is concerned with ordering problems in which the objective is not to find a subset of minimum total cost, as in the examples described above, but is instead to find a sequence that minimizes an incremental sum of costs, or equivalently, a weighted average of costs. 
For example, in many scheduling problems, the objective is to minimize some weighted sum of completion times of a set of jobs.
The order in which the jobs are processed may be constrained by some form of precedence constraints. 
In search problems, one might wish to minimize the expected time or cost incurred in searching for a target or targets that are hidden according to a known probability distribution, and the set of feasible searches may be restricted by some network structure. 
These problems arise in search and rescue as well as military search operations and can be interpreted as ``min sum'' versions of such problems as the spanning tree problem.
A ``min sum'' version of Set Cover called {\em Min Sum Set Cover} was introduced by \cite{FLT02, FLT04}.
Sequential testing problems also come under this framework: a set of tests (for example, medical tests, database queries or quality tests of computer chip components) must be performed in some order to minimize an expected cost (of forming a diagnosis, of determining whether the query is satisfied or of checking whether the component meets certain quality standards).

We unify problems of this type by introducing a new, very general problem formulation, which we call the {\em Min Sum Ordering Problem} or {\em MSOP}. Let $V$ be a finite set of cardinality $n$ and let $f,g:\mathcal{F} \rightarrow \mathbb{R} $ be a {\em cost function} and a {\em weight function}, respectively, defined on some family of subsets $\mathcal{F} \subseteq 2^V$ that contains $\emptyset$ and $V$ (where we use the symbol $\subseteq$ to denote ``is a subset of or equal to'' and $\subset$ to denote strict inclusion). 
Further, $f$ and $g$ are monotone non-decreasing with respect to set inclusion and $f(\emptyset)=g(\emptyset)=0$.
In our applications, the set $\mathcal{F}$ is typically implicitly defined by the problem setting rather than being part of the input. We assume that $f$ and $g$ are given by value oracles.
We define an {\em $\mathcal{F}$-chain} to be a sequence of subsets $\mathcal{S}=(S_j)_{j=0}^k$ for some $k \in \{1,\ldots,n\}$ such that $\emptyset = S_0 \subset S_1 \subset \ldots \subset S_k = V$ and $S_j \in \mathcal{F}$ for each $j$. When there is no ambiguity, we will simply refer to an $\mathcal{F}$-chain as a {\em chain}. 

Then the MSOP is to minimize
\begin{equation}
C_{f,g}(\mathcal{S}) \equiv \sum_{j=1}^k f(S_j)(g(S_j)-g(S_{j-1})), \label{eq:MSOP}
\end{equation}
over all chains $\mathcal{S}=\{S_j\}_{j=0}^k$, for any $k \in \{1,\ldots,n\}$. We emphasize that the minimization here is over chains of different lengths and the parameter $k$ is not an input to the problem.

If $\mathcal{S}$ minimizes $C_{f,g}(\mathcal{S})$, we say it is {\em optimal} and if $C_{f,g}(\mathcal{S})$ is at most a factor $\alpha \ge 1$ times the optimal value of the objective, we say $\mathcal{S}$ is an $\alpha$-approximation.

If $\mathcal{F}$ contains {\em all} subsets of $V$, then  the problem is equivalent to minimizing over all permutations of $V$ (see Lemma~\ref{lem:subchain}).  Here the subsets in a chain correspond to the elements picked ``so far'' by the permutation. By setting the problem up in the more general way, in terms of maximizing over chains rather than permutations, we ensure that the model is general enough to incorporate the intricacies of precedence constraints or restrictions due to a network structure, for example.

The applications we are interested in include problems of minimizing over a given {\em subset} of the set of all permutations of $V$. 
To be more precise, let $\sigma:V \rightarrow V$ be a permutation of $V$, and for $j = 1,\ldots,n$, let $S^{\sigma}_j$ be the union of the first $j$ elements of $V$ under the permutation $\sigma$. 
Then for a set $\Sigma$ of permutations of $V$ and given functions $f$ and $g$, we wish to solve
\begin{align} \label{eq:perm}
\min_{\sigma \in \Sigma}~ \sum_{j=1}^{n} f(S^{\sigma}_j)(g(S^{\sigma}_j)-g(S^{\sigma}_{j-1})).
\end{align}
We refer to this problem as the {\em Min Sum Permutation Problem} or {\em MSPP}. 
Neither one of MSOP nor MSPP can be reduced to the other, but we will show in Section~\ref{sec:main} that as long as $\Sigma$ satisfies some technical condition, MSPP can be regarded as a special case of MSOP. 
Indeed, let $\mathcal{F}^{\Sigma}$ consist of all {\em initial sets} of all $\sigma \in \Sigma$ (that is, $S	\in \mathcal{F}^\Sigma$ if and only if $S=S_j^\sigma$ for some $j$ and some permutation $\sigma \in \Sigma$).  
Suppose we have an $\alpha$-approximate solution $\mathcal{S}$ to MSOP with $\mathcal{F}=\mathcal{F}^\Sigma$. 
Then any permutation $\sigma$ such that every subset in $\mathcal{S}$ is some initial set of $\sigma$ will be an $\alpha$-approximate solution to MSPP. 
This observation is easily verified and we postpone its justification until Section~\ref{sec:main}. 

We choose to write this paper in terms of MSOP rather than MSPP for two main reasons. Firstly, the setting allows us to prove the strongest form of our main results. Secondly, working with chains rather than permutations allows the greedy algorithm we will analyze to consider the ``bigger picture'' by recursively adding {\em subsets} of elements rather than adding elements one by one.

The idea of minimizing over chains is novel, but MSPP has been studied previously in some other special cases. In particular, \cite{Pisaruk} considered the problem in the case that $f$ is submodular and $g$ is supermodular, giving a 2-approximation algorithm. (The function $f$ is submodular if and only if $f(S \cup T) + f(S \cap T) \le f(S) + f(T)$ for all $S,T$; the function $g$ is supermodular if $f(S \cup T) + f(S \cap T) \ge f(S) + f(T)$ for all $S,T$.)
This special case was also considered more recently in \cite{FLV19}, where the problem was introduced as the {\em submodular search problem}, and previous to that, \cite{FKR17} considered the case of $f$ submodular and $g$ modular. 

If $g$ is the cardinality function $g(S)=|S|$ and the sets $S_j$ increase by one element in each step, then the sum in~(\ref{eq:MSOP}) reduces to $\sum_{j=1}^k f(S_j)$. 
This special case of MSPP was considered in~\cite{ITT12}, for various classes of functions $f$. 
In particular, a 4-approximation algorithm was obtained in the case that $f$ is supermodular. 
We discuss this in more detail in Subsection~\ref{sec:MSSC}, in particular in reference to MSSC and its generalizations. A more general 4-approximation had already been proved by \cite{DBLP:conf/nips/StreeterG08} in the case of $f$ supermodular and $g$ modular.

MSOP is more general than the problems of \cite{ITT12} and \cite{Pisaruk} in two ways. 
Firstly, we make weaker assumptions about the form of the cost and weight functions and our main result requires only that the cost function is subadditive. A cost function $f$ is subadditive if and only if $f(S \cup T) \le f(S) + f(T)$ for all disjoint sets $S,T \in \mathcal{F}$.\footnote{Some definitions of subadditive set functions in the literature require that $f(S \cup T) \le f(S) + f(T)$ for non-disjoint $S$ and $T$ as well.  When $f$ is non-decreasing with respect to set inclusion, as it is in our case, this is an equivalent definition.}
Subadditivity is a more general concept than submodularity.
Secondly, the aforementioned works take the approach of minimizing over all permutations of $V$, in contrast to our approach of minimizing over chains. 

Since MSOP generalizes several NP-hard problems, MSOP is NP-hard itself, so we consider approximation algorithms for the problem.
An important concept in our analysis is that of the {\em density} $\rho(S)$ of a subset $S$ of $V$, defined by $\rho(S)=g(S)/f(S)$ (for $f(S) \neq 0$).
We will define a simple greedy algorithm for MSOP, which we now briefly describe (see Section~\ref{sec:main} for a more precise description). 
The algorithm constructs a chain by iteratively choosing the $(j+1)$th subset $S_{j+1}$ in the chain in such a way as to maximize the marginal density $(g(S_{j+1})-g(S_j))/(f(S_{j+1})-f(S_j))$. 
We refer to this problem of finding the next element of the chain as the {\em maximum density problem}.
If the maximum density problem cannot be solved in polynomial time, but we can approximate it in polynomial time within a factor $\alpha \ge 1$ then we call a chain produced by such an approximation an {\em $\alpha$-greedy} chain. 
We will prove the following in Section~\ref{sec:main}.

\begin{theorem} \label{thm:4-approxforward}
Suppose $f$ is subadditive and $\mathcal{F}$ is closed under union. Then for any $\alpha \ge 1$, an $\alpha$-greedy chain is a $4 \alpha$-approximation for an optimal chain for MSOP.
\end{theorem}

Later, in Section~\ref{sec:dual}, we consider a ``backward'' version of our greedy algorithm.
Instead of starting with the empty set of elements and adding elements in each greedy step to form successively larger sets, the backward greedy algorithm begins with $V$ and  removes elements in each greedy step to form successively smaller sets.
A backward greedy approach was previously used by \cite{ITT12} in their work on some special cases of MSPP.
We introduce a dual version of the MSOP problem, analogous to the dual problem introduced in ~\cite{FLV19}, and prove a result similar to Theorem~\ref{thm:4-approxforward} for the backward greedy algorithm and the dual MSOP problem.

The proof of Theorem~\ref{thm:4-approxforward} is inspired by the elegant proof in \cite{FLT04} of the $4$-approximation algorithm for the problem {\em Min Sum Set Cover} ({\em MSSC}). This is the problem of ordering a ground set $V$ to minimize the sum of ``covering times'' of a given collection of subsets of $V$, where the covering time of a subset is the earliest position in the ordering of any element of that subset. 
The proof uses the idea of representing the cost of the ordering produced by the greedy algorithm and that of an optimal ordering by two histograms, and showing that when the first histogram is shrunk by a factor of two in the horizontal and vertical directions, it fits in the second histogram. 
The proof idea is generalized in \cite{DBLP:conf/nips/StreeterG08}, who proved a 4-approximation result for a class of problems that includes some special cases of MSOP, including MSSC.
A different generalization of MSSC is given in \cite{ITT12} to prove the $4$-approximation for one case of the {\em Minimum Linear Ordering Problem}. 
More recently, a similar proof was used in \cite{hermans2021exact} to establish an $8$-approximation for the {\em expanding search} problem, and in \cite{HS20arxiv} to obtain a 4-approximation for \emph{bipartite OR-scheduling}. 

While the last three works cited all use a similar proof method, the proof is somewhat different in each case and none of these results directly implies another.
The similarity of the proofs strongly suggests that some deeper result is behind all of these problems. 
We confirm here that this is indeed the case by showing that MSOP generalizes each of them and Theorem~\ref{thm:4-approxforward} generalizes their respective approximation results.

In Section~\ref{sec:main} we prove Theorem~\ref{thm:4-approxforward}, using a variation of the proof originally devised by \cite{FLT04}. 
The main difference from the original proof stems from the fact that our algorithm does not greedily pick elements of $V$ one by one, but rather greedily picks subsets in the chain. 
Also, we do not optimize over permutations but over chains.

In Section~\ref{sec:applications}, we will show that Theorem~\ref{thm:4-approxforward} can be used to recover a number of known results relating to search theory and variants of MSSC. We go on to apply our results to new problems. 
In particular, we consider scheduling problems with {\em OR-precedence constraints} in Section~\ref{sec:OR-precedence}, where the set of jobs to be processed is represented by the vertices of a directed acyclic graph (DAG), and a job can only be processed after at least one of its predecessors in the DAG has been processed. 
We use Theorem~\ref{thm:4-approxforward} to show that there is a polynomial time 4-approximation algorithm for the problem of minimizing the sum of the weighted completion times of a set of jobs that must be scheduled so as to respect some OR-precedence constraints given by a DAG that is in the form of an {\em inforest} or, more generally, a {\em multitree} (where inforests and multitrees will be defined in Section~\ref{sec:OR-precedence}). We also give a 4-approximation algorithm for a version of MSSC with OR-precedence constraints in the form of an inforest.
Finally, in Section~\ref{sec:rofpf} we give an 8-approximation algorithm for minimizing the expected cost of non-adaptively evaluating a Boolean read-once formula (AND/OR tree), assuming independent tests. In Section~\ref{sec:dual} we introduce the dual problem, which leads to further approximation results, and in Section~\ref{sec:conclusion} we indicate directions for future work.


\section{Approximating MSOP} \label{sec:main}

In this section, we first prove our main result in Subsection~\ref{sec:proof}. In Subsection~\ref{sec:perm} we then justify the observation made in the introduction that provided $\Sigma$ satisfies some technical condition, MSPP is really a special case of MSOP. 

\subsection{Proof of Main Result}
\label{sec:proof}

For a set $A \in \mathcal{F}$, we write $f_A$ for the function on $\mathcal{F}$ given by $f_A(S)=f(S)-f(A)$, and similarly for $g_A$. 
For $f_A(S) \neq 0$, let $\rho_A(S)=g_A(S)/f_A(S)$ be the {\em marginal density} of $S$ (with respect to $A$); if $f_A(S)=0$ we set $\rho_A(S)=\infty$. If $A =\emptyset$, we drop the subscript from $\rho$ and simply refer to $\rho(S)$ as the density of $S$.

We consider a greedy algorithm for MSOP. For $\alpha \ge 1$, we call an $\mathcal{F}$-chain $\mathcal{S}=(S_j)_{j=0}^k$ an {\em $\alpha$-greedy chain} if 
\[
\rho_{S_j}(S_{j+1}) \ge \frac{1}{\alpha} \max_{\{T \in \mathcal{F}: S_{j} \subseteq T\}} \rho_{S_j}(T),
\]
for all $j=0,\ldots, k-1$. If $\alpha=1$, an $\alpha$-greedy chain is simply one for which $S_{j+1}$ has maximum marginal density with respect to $S_j$ for each $j=0,\ldots,k-1$.

We now prove Theorem~\ref{thm:4-approxforward}, which generalizes both the result and the proof in \cite{FLT04}.

\begin{figure}[htb!]
\centering
  \begin{subfigure}{0.8\textwidth}
    \includegraphics[width=1\textwidth]{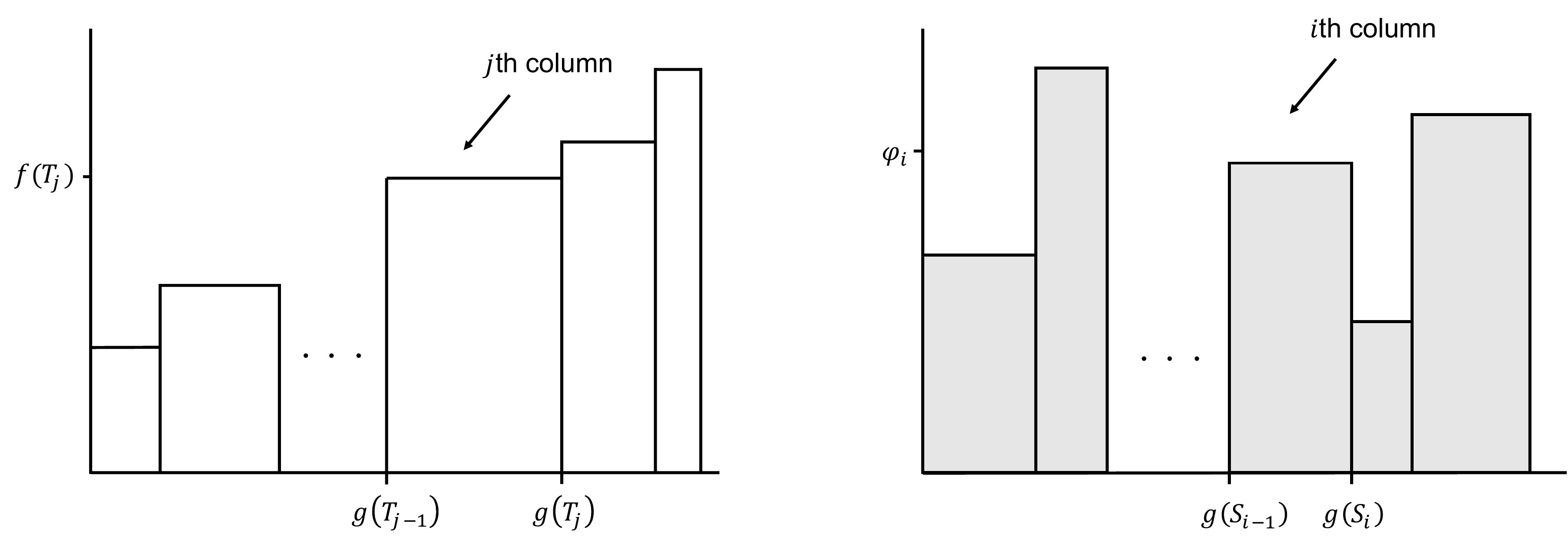}
    \caption{}
	\label{fig:hist1}
	\end{subfigure}\\
	  \begin{subfigure}{0.33\textwidth}
	  \centering
    \includegraphics[width=1\textwidth]{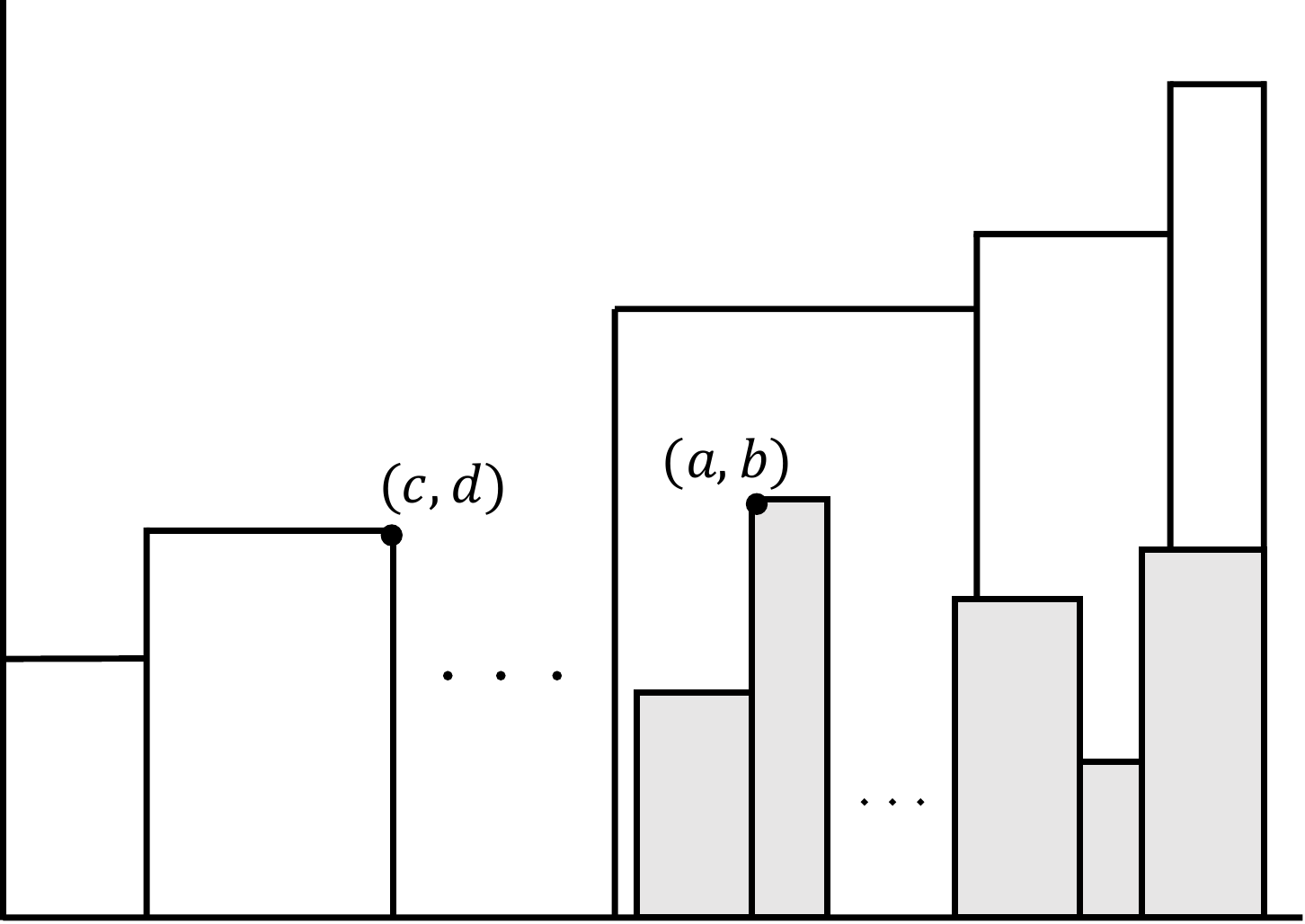}
    \caption{}
	\label{fig:hist2}
	\end{subfigure}
	\label{fig:hist}
	\caption{(a) Two histograms with total area $C_{f,g}(\mathcal{T})$ (left) and $C_{f,g}(\mathcal{S})$ (right); (b) the shrunken version of the second histogram in the first histogram.}
\end{figure}

\textit{Proof of Theorem~\ref{thm:4-approxforward}.}
Let $\mathcal{T}=(T_j)_{j=0}^\ell$ be an optimal chain and let $\mathcal{S}=(S_i)_{i=0}^k$ be an $\alpha$-greedy chain. We first construct a histogram with $\ell$ columns, the area under which is equal to $C_{f,g}(\mathcal{T})$. The base of the $j$th column of the histogram is the interval from $g(T_{j-1})$ to $g(T_j)$ and its height is $f(T_j)$. Thus, the total area under the histogram is equal to $C_{f,g}(\mathcal{T})$. The histogram is depicted in the top left of Figure~\ref{fig:hist1}.

Next, we construct a second histogram with $k$ columns, the area under which is equal to $C_{f,g}(\mathcal{S})$. Let
$\rho_i = \rho_{S_{i-1}}(S_i)$ and let $\varphi_i = \rho_i^{-1}(g(V)-g(S_{i-1}))$ (if $\rho_i=\infty$, set $\rho_i^{-1}=0$). The base of the $i$th column of this histogram is the interval from $g(S_{i-1})$ to $g(S_i)$ and its height is $\varphi_i$.  
Thus the total area $A$ under this histogram is
\begin{align}
A&= \sum_{i=1}^k \varphi_i(g(S_i)-g(S_{i-1})) \nonumber \\
&= 
\sum_{i=1}^k (g(V)-g(S_{i-1}))(f(S_i)-f(S_{i-1})) \label{eq:dual} \\
&= g(V) \sum_{i=1}^k (f(S_i)-f(S_{i-1})) -  \sum_{i=1}^k g(S_{i-1}) (f(S_i)-f(S_{i-1})) \nonumber
\end{align}

The first sum on the right-hand side above is telescopic and equal to $f(V)-f(\emptyset) = f(V)$. Rearranging the second sum, we obtain
\begin{align*}
    A &= g(V) f(V) - g(S_{k-1}) f(V) + \sum_{i=1}^{k-1} f(S_i)(g(S_i)-g(S_{i-1})) \\
    & = \sum_{i=1}^{k} f(S_i)(g(S_i)-g(S_{i-1})) \\  &= C_{f,g}(\mathcal{S}).
\end{align*}

The second histogram is depicted in the top right of Figure~\ref{fig:hist1}. Note that the heights of the columns in the first histogram, from left to right, are non-decreasing.

Now shrink the second histogram 
by a factor of $2\alpha$ in the vertical direction, and a factor of $2$ 
in the horizontal direction, and move it to the right so it is flush with the right end $g(V)$, as depicted in Figure~\ref{fig:hist2}. This results in point $(x,y)$ being mapped to  $(\frac{g(V)+x}{2},\frac{y}{2\alpha})$. 
The distance of this latter point from the right end is 
$(g(V)-x)/2$.

We now show that the shrunken (and shifted) histogram is contained in the first histogram, from which it follows that
$C_{f,g}(\mathcal{S}) \leq 4 \alpha C_{f,g}(\mathcal{T})$, proving the theorem.
To show that the shrunken histogram is contained in the first histogram, it is sufficient to show that if $(a,b)$ is the top left point of some column $i$ in the shrunken histogram, and $(c,d)$ is the top right point of some column $j$ in the first histogram, then $d < b$ implies that $c < a$. 
Here $(a,b)=\left(\frac{g(V) + g(S_{i-1})}{2},\frac{\varphi_i}{2\alpha}\right)$ and $(c,d)=(g(T_j),f(T_j))$. 

So assume $d < b$, or equivalently

\begin{equation}
\label{wehave}
f(T_j) < \frac{\rho_i^{-1}(g(V) - g(S_{i-1}))}{2\alpha}.
\end{equation}

Let $a'$ be the distance of $(a,b)$ from the right boundary, that is, 
$a' = (g(V)-g(S_{i-1}))/2$ and let $c'$ be the distance of $(c,d)$ from the right boundary, that is, $c' = g(V) - g(T_j)$.  We want to show that $d < b$ implies $c < a$, or equivalently that $a' < c'$.  So we want to show that

\begin{align}
\label{wewant}
\frac{g(V)-g(S_{i-1})}{2} < g(V) - g(T_j).
\end{align}

Since $\mathcal{F}$  is closed under union,
$S_{i-1} \cup T_j  \in \mathcal{F}$.
We will use the fact that, because $\mathcal{S}$ is $\alpha$-greedy,

\begin{equation}
\label{greedyrate}
 \rho_i = \rho_{S_{i-1}}(S_i) \geq \frac{1}{\alpha}\rho_{S_{i-1}}(S_{i-1} \cup T_j) = 
\frac{g(S_{i-1} \cup T_j) - g(S_{i-1})}{\alpha(f(S_{i-1} \cup T_j) - f(S_{i-1}))} .
\end{equation}
 
Because $f$ is subadditive and non-decreasing, 
\[
f(S_{i-1} \cup T_j) - f(S_{i-1}) \leq f(S_{i-1} \cup T_j) - f(S_{i-1} \setminus T_j) \leq f(T_j).
\] 
Combining that fact with (\ref{greedyrate}) yields
\begin{align*}
\alpha f(T_j) &\geq \rho_i^{-1}(g(S_{i-1} \cup T_j) - g(S_{i-1})) \\
 &=\rho_i^{-1}((g(V) - g(S_{i-1})) - (g(V) - g(S_{i-1} \cup T_j)))
\end{align*}

Using our assumption in (\ref{wehave}), we thus get

\begin{align*}
\frac{\rho_i^{-1}(g(V) - g(S_{i-1})) }{2} & > \rho_i^{-1}((g(V) - g(S_{i-1})) - (g(V) - g(S_{i-1} \cup T_j))) \\
& \ge \rho_i^{-1}((g(V) - g(S_{i-1})) - (g(V) - g( T_j))),
\end{align*}
where the second inequality follows from the fact $g$ is non-decreasing. Rearranging gives~(\ref{wewant}).  
$\Box$

Observe that if $\mathcal{F}=2^V$ and $f$ is supermodular and $g$ is submodular, then for $S_j \subseteq T$,
\[
\rho_{S_j}(T) \le \frac{\sum_{v \in T \setminus S_j} g(S_j\cup\{v\})-g(S_j)}{\sum_{v \in T \setminus S_j} f(S_j \cup\{v\})-f(S_j)} \le \max_{v \in T \setminus S_j} \rho_{S_j}(S_j \cup \{v\}).
\]
Hence, a $1$-greedy chain can be obtained in polynomial time by adding singletons one-by-one. Suppose $f$ is not just supermodular but also modular. Then $f$ is also subadditive so, by Theorem~\ref{thm:4-approxforward}, there is a polynomial time $4$-approximation algorithm. We summarize this observation below.

\begin{corollary} \label{cor:4approx}
Suppose $\mathcal{F}=2^V$. If $f$ is modular and $g$ is submodular then a $1$-greedy chain can be constructed in polynomial time and there exists a polynomial time $4$-approximation algorithm for MSOP.
\end{corollary}

As discussed in Subsection~\ref{sec:MSSC}, the problem MSSC and, more generally, pipelined set cover (a generalization of MSSC with costs on the elements of $V$ and weights on the sets that must be covered) are special cases of MSOP where $g$ is submodular and $f$ is modular. Thus the $4$-approximation algorithms for these problems follow from Corollary~\ref{cor:4approx}.

\subsection{Minimizing Over Permutations}
\label{sec:perm}

We now turn to the problem MSPP, given in~(\ref{eq:perm}), where we wish to minimize a weighted sum over a set of permutations $\Sigma$. Recall that $\mathcal{F}^\Sigma$ consists of all subsets of $V$ that are initial sets of some permutation $\sigma \in \Sigma$. We will show that provided $\Sigma$ satisfies a certain technical condition, solving MSPP for $\Sigma$ is equivalent to solving the corresponding MSOP problem (with the same $f$ and $g$) for $\mathcal{F}=\mathcal{F}^\Sigma$.

Suppose $\mathcal{F}$ is some family of subsets of $V$, and suppose $\mathcal{S} \equiv (S_j)_{j=0}^{k}$ is an $\mathcal{F}$-chain. Then if $0=j_0 \le j_1 \le \ldots \le j_{\ell}=k$, we say $\mathcal{S}' \equiv (S_{j_i})_{i = 0}^{\ell}$ is a {\em subchain} of $\mathcal{S}$.  It is intuitively clear that if $\mathcal{S}'$ is a subchain of $\mathcal{S}$ then $C_{f,g}(\mathcal{S}') \ge C_{f,g}(\mathcal{S})$, since $\mathcal{S}'$ advances in ``bigger steps''. We formalize this in the next lemma.

\begin{lemma} \label{lem:subchain}
Suppose $\mathcal{S'} \equiv (S_{j_i})_{i = 0}^{\ell}$ is a subchain of the $\mathcal{F}$-chain $\mathcal{S} \equiv (S_j)_{j=0}^{k}$. Then \begin{enumerate}
    \item[(i)] $C_{f,g}(\mathcal{S'}) \ge C_{f,g}(\mathcal{S})$ and
    \item[(ii)] if $\mathcal{S'}$ approximates MSOP by a factor of $\alpha \ge 1$ then so does $\mathcal{S}$.
    \end{enumerate}
\end{lemma}
\textit{Proof.} We perform the following calculation.
\begin{align*}
    C_{f,g}(\mathcal{S'}) &\equiv \sum_{i=1}^{\ell} f(S_{j_i})(g(S_{j_i})-g(S_{j_{i-1}})) \\
    &= \sum_{i=1}^{\ell} f(S_{j_i}) \sum_{j=j_{i-1}+1}^{j_i} (g(S_j) - g(S_{j-1}) ) \\
    &\ge \sum_{i=1}^{\ell} \sum_{j=j_{i-1}+1}^{j_i} f(S_{j}) (g(S_j) - g(S_{j-1}) ) \\
    &= \sum_{j=1}^k f(S_j) (g(S_j) - g(S_{j-1})) \equiv C_{f,g}(\mathcal{S}),
\end{align*}
where the inequality above follows from the monotonicity of $f$ and $g$. Part (ii) of the lemma follows immediately.  
$\Box$

Suppose now that $\Sigma$ is a set of permutations of $V$. 
If $\mathcal{S}$ is an $\mathcal{F}^{\Sigma}$-chain and  $\sigma \in \Sigma$ is a permutation, such that each element of $\mathcal{S}$ is an initial set of $\sigma$, then we say $\sigma$ is {\em consistent} with $\mathcal{S}$. If every $\mathcal{F}^\Sigma$-chain is consistent with some permutation in $\Sigma$ then we say $\Sigma$ is {\em well-founded}.

\begin{lemma} \label{lem:perms}
Suppose $\Sigma$ is a set of permutations of $V$ and that $\Sigma$ is well-founded. If there exists a polynomial time $\alpha$-approximation algorithm for MSOP with $\mathcal{F}=\mathcal{F}^\Sigma$ for some $\alpha \ge 1$ then there exists a polynomial time $\alpha$-approximation algorithm for MSPP with $\Sigma$.
\end{lemma}
\textit{Proof.}
This follows immediately from Lemma~\ref{lem:subchain}, part (ii). Indeed, first note that every $\mathcal{F}^\Sigma$-chain is consistent with a permutation whose objective value is no greater and every permutation is consistent with an $\mathcal{F}^\Sigma$-chain with the same objective value. It follows that the objective value of any optimal solution to MSOP is equal to the objective value of any optimal solution to MSPP.

Now suppose that $\mathcal{S'}$ is an $\alpha$-approximate $\mathcal{F}^\Sigma$-chain and that $\sigma$ is consistent with $\mathcal{S'}$. Let $\mathcal{S}$ be the chain consisting of all the initial sets of $\sigma$. 
Then $\mathcal{S'}$ is a subchain of $\mathcal{S}$, so $\mathcal{S}$ is an $\alpha$-approximation for MSOP. Equivalently, $\sigma$ is an $\alpha$-approximation for MSPP.   
$\Box$

Note that Lemma~\ref{lem:perms} trivially holds in the opposite direction. Indeed, if there exists a polynomial time $\alpha$-approximation for MSPP, then this also provides a polynomial time $\alpha$-approximation for MSOP with $\mathcal{F}=\mathcal{F}^\Sigma$, since every permutation $\sigma$ defines a feasible $\mathcal{F}^\Sigma$-chain with the same objective function value.

It is easy to think of examples of $\Sigma$ that are not well-founded. For example, if $V=\{1,2,3\}$ and $\Sigma$ contains only the permutations $(1,2,3)$ and $(3,1,2)$, then the $\mathcal{F}^\Sigma$-chain $\{\{1\},\{1,3\},\{1,2,3\}\}$ is not consistent with either of the two permutations, so $\Sigma$ is not well-founded.

However, for all the examples we consider in this paper, the set of permutations is well-founded. This is easy to check by using the following sufficient condition. 

For two permutations $\sigma$ and $\tau$ of $V$, let $\pi_j(\sigma,\tau)$ be the permutation that follows $\sigma$ for the first $j$ elements, then chooses the remaining elements of $V$ in the order specified by $\tau$, for each $1 \le j \le n$. For example, if $V=\{1,2,3,4,5\}$, and $\sigma$ and $\tau$ are given by $(3,1,5,2,4)$ and $(4,5,1,2,3)$, respectively then $\pi_2(\sigma,\tau)$ is given by $(3,1,4,5,2)$ and $\pi_3(\sigma,\tau)$ is given by $(3,1,5,4,2)$. We call this operation {\em splicing}.

If $\Sigma$ is a set of permutations for which $\pi_j(\sigma,\tau) \in \Sigma$ for any $\sigma, \tau \in \Sigma$ and $1 \le j \le n$, then we say $\Sigma$ is {\em closed under splicing}. 

\begin{lemma} \label{lem:closure}
Let $\Sigma$ be a set of permutations of $V$. If $\Sigma$ is closed under splicing then it is well-founded and $\mathcal{F}^\Sigma$ is closed under union.
\end{lemma}
\textit{Proof.}
Suppose $\Sigma$ is closed under splicing. Let $\mathcal{S}=(S_j)_{j=1}^k$ be an $\mathcal{F}^\Sigma$-chain. We will show that there is some permutation contained in $\Sigma$ that is consistent with $\mathcal{S}$. Let $\sigma_j$ be a permutation in $\Sigma$ that is consistent with $S_j$ for $j=1,\ldots,k$. We set $\tau_1=\sigma_1$ and for $j=2,3,\ldots,k$, we recursively define $\tau_j=\pi_{|S_{j-1}|}(\tau_{j-1},\sigma_j)$,  which is contained in $\Sigma$, by induction on $j$ and since $\Sigma$ is closed under splicing. Also,  $S_1$ is an initial set of $\tau_1$, and, by definition of $\pi_{|S_{j-1}|}(\tau_{j-1},\sigma_j)$ and by induction on $j$, each of $S_1,\ldots,S_j$ are initial sets of $\tau_j$ for $j \ge 2$. Therefore, $\tau_k$ is consistent with $\mathcal{S}$, so $\Sigma$ is well-founded.

To see that $\mathcal{F}^\Sigma$ is closed under union, let $S$ and $T$ be elements of $\mathcal{F}^\Sigma$. Then they are initial sets of some permutations $\sigma$ and $\tau$ in $\Sigma$, so $S \cup T$ is an initial set of $\pi_{|S|}(\sigma,\tau)$, which lies in $\Sigma$, since $\Sigma$ is closed under splicing. Therefore $S \cup T \in \mathcal{F}^\Sigma$.  
$\Box$

We observe that for the expanding search problem, if we take $\Sigma$ to be the set of expanding searches, then it is easy to check that $\Sigma$ is closed under splicing and therefore well-founded, by Lemma~\ref{lem:closure}. Furthermore, $\mathcal{F}^\Sigma$ is closed under union.
Similarly, for both AND-precedence constraints and OR-precedence constraints, the set of feasible orderings is closed under splicing and therefore well founded. 
If $\Sigma$ consists of all permutations of $V$, as in Boolean function evaluation, then $\Sigma$ is trivially well-founded. 
It follows from Lemma~\ref{lem:perms} that for these problems, if we can find a solution (or approximate solution) $\mathcal{S}$ to MSOP with $\mathcal{F}=\mathcal{F}^{\Sigma}$, then we can recover a solution (or approximate solution) to the original problem by taking any permutation that is consistent with $\mathcal{S}$. Since in each case $\mathcal{F}^\Sigma$ is closed under union, we only need $f$ to be subadditive to apply Theorem~\ref{thm:4-approxforward}.

\section{Special cases of MSOP} \label{sec:applications}

In this section we describe some special cases of MSOP, including those for which our results imply the existence of approximation algorithms that are already known in the literature. We will begin in Subsection~\ref{sec:search} by discussing problems in search theory and scheduling, including the recent 8-approximation result of \cite{hermans2021exact} for expanding search. Then, in Subsection~\ref{sec:MSSC}, we will describe how a number of approximation results for Min Sum Set Cover and its generalizations follow from our results.

\subsection{Search Theory and AND-Scheduling} \label{sec:search}

The expanding search problem was introduced in \cite{AL13}, and independently in \cite{AP12} under different nomenclature. 
A connected graph $G=(V,E)$ is given, and each edge $e\in E$ has a cost $c_e$. 
A target is hidden on one of the vertices of the graph according to a known probability distribution, so that the probability it is on vertex $v\in V$ is $p_v$. 
An {\em expanding search}, starting at some distinguished root $r$, is a sequence of edges $e_1,\ldots,e_{|E|}$ chosen so that $r$ is incident to $e_1$ and every edge $e_i$ ($i >1$) is incident to some previously chosen edge. 
For a given expanding search, the {\em expected search cost} of the target is the expected value of the sum of the costs of all the edges chosen up to and including the first edge that contains the target. 
The problem is to find an expanding search with minimal expected search cost. 
The problem was shown to be NP-hard in \cite{AP12}, and \cite{hermans2021exact} recently gave an 8-approximation algorithm.

To express the problem in the form of MSOP, let $\Sigma$ be the set of expanding searches and take $\mathcal{F}=\mathcal{F}^{\Sigma}$.
Then $\mathcal{F}$ is closed under union, since its elements consist of connected subgraphs of $G$ containing $r$.
For $S\in \mathcal{F}$, let $f(S)=\sum_{e \in S} c_e$ and let $g(S)$ be the sum of $p_v$ over all vertices $v$ contained in some edge of $S$. 
Then $f$ is modular, and Theorem~\ref{thm:4-approxforward} along with the results of Subsection~\ref{sec:perm} on MSPP imply that the greedy algorithm is $4 \alpha$-approximate, where $\alpha$ is the approximation ratio of the maximum density problem. This coincides with the algorithm in Theorem~2 of \cite{hermans2021exact}.

\cite{AL13} gave a solution to the expanding search problem in the case that the graph is a tree. 
In this case, the problem is equivalent to a special case of the single machine, precedence constrained scheduling problem, usually denoted $1|prec|\sum w_j C_j$, of minimizing the sum of the weighted completion times of a set of jobs. 
A partial order is given on the jobs, and a job becomes available for processing only after all of its predecessors have been processed. 
We refer to this type of precedence constraints and precedence constrained scheduling as {\em AND-precedence constraints} and {\em AND-scheduling}. 
The jobs have weights $w_j$ and processing times $p_j$, and for a given ordering, the completion time $C_j$ of a job $j$ is the sum of its processing time and all the processing times of the jobs preceding it.  
The problem is to find a feasible ordering that minimizes the weighted sum $\sum w_j C_j$ of the completion times. 
Comparing the weights and the processing times to the probabilities and the edge costs in the expanding search problem, it is easy to see that if the partial order has a tree-like structure, then the scheduling problem and the search problem are equivalent, as pointed out in \cite{FLV19}.

A polynomial time algorithm for the scheduling problem $1|prec|\sum w_j C_j$ on trees was given by~\cite{horn1972single}. \cite{Sidney} proved that any optimal schedule (for general precedence constraints) must respect what is now known as a {\em Sidney decomposition}, obtained by recursively taking subsets of jobs of maximum density. \cite{lawler1978sequencing} gave a polynomial time algorithm for the problem on series-parallel graphs, which was generalized to two-dimensional partial orders in a series of papers of \cite{CS05} and~\cite{AM09}.
\cite{CM99} and \cite{MQW03} independently showed that any schedule consistent with a Sidney decomposition is a 2-approximation.
Earlier 2-approximation algorithms were derived by \cite{Schulz} and \cite{CH99}. 
\cite{CS05} showed that all known $2$-approximations are consistent with a Sidney decomposition.
Sidney's decomposition theorem and the resulting 2-approximation algorithm was generalized to the case of MSPP with $f$ submodular and $g$ supermodular in \cite{FLV19}, where further applications to scheduling and search problems were given. 

The idea of a Sidney decomposition essentially coincides with this paper's main algorithm, but for the special cases described above, a different approach is used to prove that the algorithms are 2-approximations.

\subsection{Min Sum Set Cover and its Generalizations} \label{sec:MSSC}

Min Sum Set Cover was first introduced by~\cite{FLT02}.
An instance of MSSC consists of a finite ground set $V$ and a collection of subsets (or {\em hyperedges}) $\mathcal{E}$ of $V$.
For a given linear ordering (or permutation) $\pi : V \to [n] := \{1,\dots,n\}$ of the elements of $V$, the \emph{covering time of set} $e \in \mathcal{E}$ is the first point in time that an element contained in $e$ appears in the linear ordering, i.e., $\pi(e) := \min \{\pi(v) \, | \, v \in e \}$.
The objective is to find a linear ordering that minimizes the total sum of covering times, $\sum_{e \in \mathcal{E}} \pi(e)$.\footnote{We note that this definition of Min Sum Set Cover, given in~\cite{FLT02}, uses a ``hitting set'' formulation of the problem, in which vertices cover (hit) hyperedges, rather than vice versa.}

\cite{ITT12} introduced a generalization of MSSC called the Minimum Linear Ordering Problem, which can be regarded as the special case of MSPP where $g$ is the cardinality function. 
(In fact, \cite{ITT12} perform the summation in the opposite order from~(\ref{eq:MSOP}), but of course the objective is the same.) 
We give here a slightly different reduction of MSSC to MSPP. Taking $\mathcal{F}$ to be $2^V$, for a subset $S \in \mathcal{F}$, we define $g(S)$ to be the number of hyperedges that contain some element of $S$ and $f(S)$ to be the cardinality of $S$.
Then the total sum of covering times is given by~(\ref{eq:MSOP}). The dual of this problem (see Section~\ref{sec:dual}) corresponds to the reduction of~\cite{ITT12}.

MSSC is closely related to Minimum Color Sum (MCS), which was introduced by~\cite{KS89}, and can be shown to be a special case of MSSC (though the reduction is not of polynomial size -- see~\cite{FLT02}).
MSSC and MCS are min sum variants of the well-known Set Cover and Graph Coloring problems, respectively.
\cite{KS89} observed that MCS can be solved in linear time for trees, and~\cite{BNK98} proved that it is APX-hard already for bipartite graphs.
For general graphs,~\cite{BNBHST98} showed that a greedy algorithm is $4$-approximate for MCS. 
\cite{FLT02} observed that the greedy algorithm of~\cite{BNBHST98} applied to MSSC, which is to choose the element that is contained in the most uncovered sets next, yields a 4-approximation algorithm for MSSC.
They simplified the proof by analyzing the performance ratio via a time-indexed linear program instead of comparing the greedy solution directly to the optimum.
In the journal version of their paper,~\cite{FLT04} further simplified the proof to an elegant histogram framework, which inspired the results of this paper. They additionally proved that one cannot approximate MSSC strictly better than 4, unless $\text{P} = \text{NP}$. 

\cite{MBMW05} generalized MSSC by introducing non-negative costs $c_v$ on the elements of $V$ and non-negative weights $w_e$ on the sets in $\mathcal{E}$.
The task is to find a linear ordering $\pi$ of $V$ that minimizes the sum of weighted covering costs of the sets, $\sum_{e \in \mathcal{E}} w_e C(e)$. The covering cost of $e \in \mathcal{E}$ is defined as $C(e) := \min\{\sum_{u \in V: \pi(u) \leq \pi(v)} c_u \, | \, v \in e \in \mathcal{E}\}$ (that is, the sum of all the costs of all the elements of $V$ chosen up to and including the element that covers $e$).
This problem is known as \emph{pipelined set cover}.
A natural extension of the greedy algorithm is to pick the element $v$ that maximizes the ratio of the sum of the weights of the sets covered by $v$ and the cost of $v$.
In fact,~\cite{MBMW05} showed that this greedy algorithm is 4-approximate for pipelined set cover.

Pipelined set cover can be expressed in the form of MSOP by taking $f(S)=\sum_{v \in S} c_v$ for a subset $S \subseteq V$ and $g(S)$ to be the sum of the weights of all the subsets in $\mathcal{E}$ that contain at least one element of $S$. In this case, $g$ is submodular and $f$ is modular, and the fact that the greedy algorithm is 4-approximate follows from Theorem~\ref{thm:4-approxforward} of this paper (or more specifically, from Corollary~\ref{cor:4approx}).

Yet another generalization of MSSC is precedence-constrained MSSC.
Here, the sets are subject to AND-precedence constraints and the task is to find a linear extension of the partial order on the sets.
This problem was studied by~\cite{McCMW17}, who proposed a $4\sqrt{n}$-approximation algorithm for precedence-constrained MSSC using a similar approach to ours: first, apply a $\sqrt{n}$-approximation for finding a subset of $V$ of maximum density, and then use a histogram-type argument, which yields an additional factor of $4$. This result also follows from Theorem~\ref{thm:4-approxforward} of this paper (assuming the approximation result for the maximum density problem).

\section{OR-precedence constraints} \label{sec:OR-precedence}

We now show how our results can be applied to give new approximation algorithms for problems involving OR-precedence constraints. In Subsection~\ref{sec:applications:OR} we will define OR-scheduling and provide $4$-approximation algorithms for OR-scheduling on inforests and, more generally, on {\em multitrees}. We then consider a new OR-precedence constrained version of MSSC in Subsection~\ref{sec:applications:MSSC-OR} and show that there is also a $4$-approximation algorithm for this. 

\subsection{OR-Scheduling} \label{sec:applications:OR}

One can interpret pipelined set cover (described in Subsection~\ref{sec:MSSC}) as a single-machine scheduling problem in the following way.
There is a job $j_v$ for every element $v \in V$ with processing time $p_{j_v} = c_v$ and weight $w_{j_v} = 0$, and a job $j_e$ for every $e \in \mathcal{E}$ with processing time $p_{j_e} = 0$ and weight $w_{j_e} = w_e$.
Further, there are \emph{OR-precedence constraints} between job $j_v$ and all jobs $j_e$ with $v \in e$.
That is, job $j_e$ becomes available for processing after \emph{at least one} of its predecessors in $\{j_v \, | \, v \in e \in \mathcal{E}\}$ is completed.
Then, finding a linear ordering of $V$ that minimizes the sum of weighted covering costs is equivalent to finding a feasible single-machine schedule that minimizes the sum of weighted completion times.

Formally, OR-scheduling is defined as follows: Let $N$ be a set of jobs that are subject to precedence constraints given by a DAG $G = (N,E)$.
An edge $(i,j) \in E$ indicates that job $i$ is an OR-predecessor of $j$.
Any job $j$ with $\{i \in N \, | \, (i,j) \in E\} \not= \emptyset$ requires that at least one of its predecessors is completed before it can start.
A job without predecessors may be scheduled at any point in time.
The task is to find a feasible schedule, i.e., each job is processed non-preemptively for $p_j$ units of time, and at each point in time at most one job is processed, that minimizes the sum of weighted completion times.

To see that OR-scheduling is indeed a special case of MSOP, let $\Sigma$ be the set of feasible schedules, and let $\mathcal{F}=\mathcal{F}^\Sigma$.
In other words, $S \in \mathcal{F}$ if and only if for any job in $S$ with predecessors, at least one of its predecessors is contained in $S$ as well. Clearly, $\mathcal{F}$ is closed under union. Further, we set $f(S) = \sum_{j \in S} p_j$ and $g(S) = \sum_{j \in S} w_j$ for every set of jobs $S \subseteq N$.
With these modular functions, it is not hard to see that the sum of weighted completion times of a schedule is equal to~(\ref{eq:perm}).

Note that, for the above reduction from pipelined set cover, the set of jobs can be partitioned into $N = A \dot{\cup} B$ such that all edges in the precedence graph go from $A$ to $B$.
We call such a precedence graph \emph{bipartite}.
In a recent paper,~\cite{HS20arxiv} presented a 4-approximation algorithm for scheduling with bipartite OR-precedence constraints using an approach similar to ours.
For bipartite OR-scheduling, the maximum density sets can be computed in polynomial time, so a histogram argument yields a $4$-approximation algorithm.

Scheduling with OR-precedence constraints was previously considered in the context of AND/OR-networks, see, e.g.,~\cite{GL95,EKM03}. In this case,~\cite{EKM03} presented the best-known approximation factor, which is linear in the number of jobs, and showed that obtaining a polynomial time constant-factor approximation algorithm is NP-hard.
For the case where the AND/OR-constraints are of a similar bipartite structure as above, and no AND-constraints are within $B \times A$,~\cite{HS20waoa} obtained a $2\Delta$-approximation algorithm with $\Delta$ being the maximum number of OR-predecessors of any job in $B$.
\cite{J05} proved that minimizing the sum of weighted completion times with OR-constraints is already NP-hard for unit-processing time jobs.
\cite{HS20arxiv} strengthened this result and showed that the problem remains NP-hard even for bipartite OR-precedence constraints with unit processing times and 0/1 weights, or 0/1 processing times and unit weights.

We now consider a special case of MSOP that can be stated in terms of MSPP. Suppose the elements of $V$ are vertices of a DAG $G=(V,E)$ which represents some OR-precedence constraints. That is, a permutation $\sigma$ of $V$ is feasible if each element $v$ with a non-empty set of predecessors appears in $\sigma$ later than at least one of its direct predecessors $\mathcal{P}(v)$ (where $\mathcal{P}(v)$ is the set of all $u$ such that $(u,v)\in E$). Recall that a connected DAG $G=(V,E)$ is an {\em intree} if every vertex has at most one successor. A DAG whose connected components are intrees is an inforest.

\begin{theorem} \label{thm:intrees}
Consider an instance of MSPP for which the set of feasible permutations is derived from some OR-precedence constraints given by an inforest. Then if $f$ is modular and $g$ is submodular, there is a polynomial time $4$-approximation algorithm for the problem.
\end{theorem}

\textit{Proof.} Note $f$ is modular and hence subadditive. Also, as pointed out in Subsection~\ref{sec:perm}, the set of feasible permutations $\Sigma$ is closed under splicing and therefore, by Lemma~\ref{lem:closure}, the set $\Sigma$ is well-founded and $\mathcal{F}^\Sigma$ is closed under union. 
Hence, by Theorem~\ref{thm:4-approxforward} and Lemma~\ref{lem:perms}, it suffices to construct in polynomial time a $1$-greedy $\mathcal{F}^{\Sigma}$-chain. 

We characterize inclusion-minimal sets of maximum density, using the concept of a {\em stem}. We define a stem in $G$ to be a sequence of vertices $v_1,\ldots,v_k$ in $V$ such that $v_1$ has no predecessors and $v_{i} \in \mathcal{P}(v_{i+1})$ for all $i=1,\ldots, k-1$. We show that any inclusion-minimal subset $S_{j+1}$ of $V$ that maximizes the density $\rho_{S_{j}}(S_{j+1})$ is a stem and that we can enumerate all stems in polynomial time.
Observe that, if we remove a stem $S_j$ from the instance along with all edges incident to vertices in the stem, the graph decomposes into intrees again, since by definition of an inforest, any subgraph of an inforest is an inforest. Also $f_{S_j}$ is modular and $g_{S_j}$ is submodular.
So it suffices to consider only $S_j = \emptyset$.

Since a stem is fully specified by its starting and ending vertices, the total number of stems is $\mathcal{O}(n^2)$.
We can therefore enumerate all stems $S$ that start at a job without a predecessor, and pick the one of maximum density $\rho(S)=g(S)/f(S)$.
It remains to show that a stem of maximum density is indeed an OR-initial set of maximum density.

Let $S \in \mathcal{F}$ be an inclusion-minimal set of maximum density and suppose that $S$ is not a stem. Since $G$ is an inforest, $S$ must induce an inforest that contains at least two vertices without a predecessor.
Since every vertex has at most one successor, any vertex $v$ without a predecessor induces a unique stem from $v$ to the root of its connected component in $S$ (the root of a component being the unique vertex contained in that component that has no successors).
For such a vertex $v$, let $T_v$ be the path along the stem induced by $v$ that starts at $v$ and ends at the predecessor of the first vertex encountered whose in-degree in $S$ is greater than 1 (or ending at the root in $S$ of the stem if no such vertex exists).
Clearly, $T_v \in \mathcal{F}$, and also $\overline{T} := S \setminus T_v$ is an OR-initial set.

		
By the submodularity of $g$ and modularity of $f$, the density $\rho(S)$ satisfies
\begin{align} \label{rho:equivalence1}
\rho(S) \le \frac{g(T_v) + g(\overline{T})}{f(T_v)+f(\overline{T})} = \theta \rho(T_v) + (1-\theta)\rho(\overline{T}),
\end{align}
where $\theta = f(T_v)/(f(T_v) + f(\overline{T})) \in [0,1]$.
Hence, by the maximality of $\rho(S)$, both $T_v$ and $\overline{T}$ must have maximum density, contradicting the assumption that $S$ was an inclusion-minimal subset of maximum density.  
$\Box$

Consider the OR-scheduling problem on a DAG $G$ in the case that $G$ is an inforest. 
Since the cost function and the weight function are both modular, the next theorem follows immediately from Theorem~\ref{thm:intrees}.

\begin{theorem}\label{thm:OR:intree:4approximation}
There is a polynomial time $4$-approximation algorithm for OR-scheduling of inforests.
\end{theorem}

In fact, we derive a more general result for OR-scheduling of {\em multitrees}, introduced in \cite{furnas1994multitrees}. A DAG is called a multitree if, for every vertex, its successors form an outtree (where an outtree is a DAG such that each vertex has at most one predecessor). Equivalently, there is at most one directed path between any two vertices. Inforests are examples of multitrees.

\begin{theorem}
There is a polynomial time $4$-approximation algorithm for OR-scheduling of multitrees.
\end{theorem}
\textit{Proof.} By Theorem~\ref{thm:4-approxforward} and Lemma~\ref{lem:perms}, it is sufficient to find a $1$-greedy $\mathcal{F}$-chain, where $\mathcal{F}=\mathcal{F}^\Sigma$ and $\Sigma$ is the set of feasible schedules. 

We will show that the inclusion-minimal sets of maximum density are outtrees. This means that we can find a maximum density subset of $\mathcal{F}$ by considering each vertex $v$ with no predecessors and finding a maximum density subtree $T_v \in \mathcal{F}$ of the outtree formed by $v$ and its successors. This can be done in polynomial time using the dynamic programming algorithm of \cite{horn1972single}, for example. We then choose a subtree $T_v$ with maximum density over all vertices $v$ with no predecessors.

The proof that the inclusion-minimal sets of maximum density are outtrees is similar to the proof that the inclusion-minimal subsets of Theorem~\ref{thm:intrees} were stems, so we do not go into detail. It can be shown that if $S$ is an inclusion-minimal set of maximum density that is not an outtree, then it can be expressed as the disjoint union of an outtree $T$ and another set in $\mathcal{F}$. By an identical calculation as in~(\ref{rho:equivalence1}), the set $T$ must have maximum density, contradicting $S$ being inclusion-minimal. This completes the proof. 
$\Box$

It is worth pointing out that approximating the problem of minimizing the sum of weighted completion times for OR-scheduling appears to be harder than the analogous problem for AND-scheduling in the following sense.
As discussed in Subsection~\ref{sec:search}, for AND-scheduling there is a polynomial time algorithm for series-parallel DAGs and polynomial time 2-approximation algorithms for arbitrary DAGs, whereas for OR-scheduling, we have given polynomial time 4-approximation algorithms for inforests and, more generally, multitrees. 
It is not possible that better approximations exist for OR-scheduling on multitrees (or even for bipartite graphs) unless $\text{P} = \text{NP}$, since the same is true of MSSC \citep{FLT04}, which is a special case of OR-scheduling on bipartite graphs. Of course, for outtrees, OR-scheduling and AND-scheduling are equivalent, but there are no polynomial time algorithms known for OR-scheduling on any other classes of DAGS.

\subsection{OR-Precedence Constrained MSSC and Pipelined Set Cover} \label{sec:applications:MSSC-OR}

Consider a variation of MSSC in which the order that the elements of $V$ are chosen must be consistent with some OR-precedence constraints given by a DAG $G$. 
As in pipelined set cover, we additionally assume that there is a non-negative cost $c_v$ for each vertex $v$ and a non-negative weight $w_e$ for each hyperedge $e \in \mathcal{E}$, and the objective is to minimize the weighted sum of covering times of the edges. 
As for OR-scheduling, we take $\mathcal{F}$ to be the collection of OR-initial sets of $G$ and as for pipelined set cover, we take $f(S)=\sum_{v \in S} c_v$ and $g(S)$ to be the sum of the weights of all hyperedges in $\mathcal{E}$ that contain at least one element of $S$, for $S \in \mathcal{F}$. 
Then $f$ is modular and $g$ is submodular, so we can again apply Theorem~\ref{thm:intrees}.

\begin{theorem} \label{thm:OR-MSSC} 
There is a polynomial time $4$-approximation algorithm for pipelined set cover with OR-precedence constraints that take the form of an inforest.
\end{theorem}

To see that Theorem~\ref{thm:OR-MSSC} is a generalization of Theorem~\ref{thm:OR:intree:4approximation}, simply observe that if the set of hyperedges $\mathcal{E}$ consists of all the singletons of $V$, then pipelined set cover with OR-precedence constraints is equivalent to OR-scheduling.

\section{Evaluation of Read-Once Formulas}
\label{sec:rofpf}

In this section we give an $8$-approximation algorithm for a non-adaptive version of a Boolean function evaluation problem, involving the evaluation of a read-once formula on an initially unknown input in a stochastic setting.  
We call this the {\em na-ROF evaluation problem}.  

A read-once formula, also called an AND/OR tree, is a rooted tree with the following properties.  Each internal node of the tree is labeled either OR or AND (corresponding to OR or AND gates).  The leaves of the tree are labeled with  Boolean variables $x_1, \ldots, x_n$, where $n$ is the number of leaves, with each $x_i$ appearing in exactly one leaf.  Given a Boolean assignment to the variables in the leaves, the value of the formula on that assignment is defined recursively in the usual way: the value of a leaf labeled $x_i$ is the assignment to $x_i$, and the value of a tree whose root is labeled OR (respectively, AND) is the Boolean OR (respectively, AND) of the values of the subtrees of that root.  Read-once formulas are equivalent to series-parallel systems (cf.~\cite{Unyulurt04}).

\begin{figure}[t]
\centering

{
    \begin{tikzpicture}[-,>=stealth', level/.style={sibling distance = 4cm/#1, level distance = 1.5cm}, scale=0.6,transform shape]
    \node [treenode] {$\wedge$}
    child
    {
        node [treenode] {$x_1$}
    }
    child[treenode]
    {
        node [treenode] {$x_2$}   
    }
    child
    {
        node [treenode] {$\vee$} 
        child
        {
            node [treenode] {$\wedge$} 
            child
            {
                node [treenode] {$x_3$} 
            }
            child
            {
                node [treenode] {$x_4$}
            }
         } 
         child
        {
                node [treenode] {$x_5$} 
        }
    }
    
;
\end{tikzpicture}
}
\caption{Read-once formula.  \label{fig:rof}}
\end{figure}
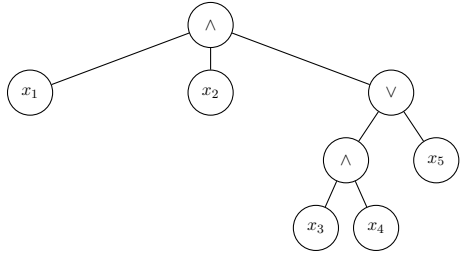

An example read-once formula, corresponding to the expression
$\phi(x_1,x_2,x_3,x_4,x_5)$ $=$ $x_1 \wedge x_2 \wedge ((x_3 \wedge x_4) \vee x_5)$, is shown in Figure~\ref{fig:rof}.  

In the na-ROF evaluation problem,
 we are given a read-once formula $\phi(x_1,\ldots,x_n)$ that must be evaluated on an initially unknown random assignment to its input variables.  For each of the $n$ Boolean variables $x_i$, we are given a probability $p_i$, where $0 < p_i < 1$, and a positive integer cost $c_i$, 
The random assignment on which we must evaluate $\phi$ is assumed to be drawn from the product distribution defined by the $p_i$'s, that is, the joint distribution in which $p_i = P[x_i=1]$ and the $x_i$ are independent.  
The value of an $x_i$ in the random assignment can only be ascertained by performing a test,  which we
call {\em test $i$}.
Performing test $i$ incurs cost $c_i$, and its outcome is the value of $x_i$.   
Tests are performed sequentially until there is enough information to determine the value of~$\phi$.   The problem is to develop a linear ordering (permutation) of the tests,
such that performing the tests in that order, until the function value is determined,
minimizes the expected cost incurred in testing.  

For example, consider
evaluating the formula in Figure~\ref{fig:rof} in the order given by the permutation $(3,4,5,2,1)$.
Suppose the first test reveals that $x_3=0$, the second that $x_4=1$, and the third that $x_5=0$.   Then testing will stop after that third test, when it can be determined that the value of $f$ is 0.  The probability of this happening is $(1-p_3) p_4(1-p_5)$, and the incurred cost in this case is $c_3 + c_4 + c_5$.
More generally, for each prefix $R$ of $(3,4,5,2,1)$, let $P_R$ denote the probability that testing stops at the end of that prefix. Using $q_i$ to denote $1-p_i$, we have e.g., $P_{(3)} = P_{(3,4)} = 0$, $P_{(3,4,5)} = (1-p_3 p_4)q_5$ and $P_{(3,4,5,2)} = (1-P_{(3,4,5)})q_2$.   The expected cost of evaluating the formula according to the permutation $(3,4,5,2,1)$ is 
$c_3 + c_4 + c_5 + c_2(1-P_{(3,4,5)}) + c_1(1-P_{(3,4,5,2)})$.
More generally, the  expected cost associated with an arbitrary permutation of the five tests is equal  to 
$\sum_{j=1}^5 f(S_j)(g(S_j)-g(S_{j-1}))$, where $S_j$ is the set consisting of the first $j$ elements of the permutation $(3,4,5,2,1)$, $f(S_j) = \sum_{i  \in S_j}  c_i$, and $g(S_j)$ is the probability that the value of $\phi$ can be determined by performing just the tests in $S_j$.  Thus
$g(S_j)-g(S_{j-1}) = P_{R^j}$, where $R^j$ denotes the prefix consisting of the first $j$ elements of the permutation.

The na-ROF evaluation problem corresponds to the MSOP with
$\mathcal{F} = 2^V$,  where $V = \{1, \ldots, n\}$
is the set of tests, $f(S) = \sum_{i \in S} c_i$, and $g(S)$ is equal to the probability that the value of $\phi$ can be determined from the outcomes of the
tests in $S$.  
(We note that the same correspondence holds for analogous evaluation problems involving other types of formulas and functions. 
~\cite{GGHK18} studied the analogous  evaluation problem for symmetric Boolean functions.)  

In Section~\ref{sec:rofpf} we show that there is an 8-approximation algorithm for the na-ROF evaluation problem that runs in pseudo-polynomial time for general costs, and polynomial time for unit costs.  We do this by giving a dynamic programming algorithm producing a 2-approximate solution for the associated maximum density problem.  

To our knowledge, the na-ROF evaluation problem has not been previously studied.
However, an ``adaptive'' version of this evaluation problem has been studied since the 1970's, under a variety of names, including ``satisficing strategies for AND/OR trees'' (\cite{GHJM06}), ``sequential testing of series-parallel systems''(\cite{Unyulurt04}), and ``the Stochastic Boolean Function Evaluation (SBFE) problem for read-once formulas'' (\cite{DHK16}).  This version seeks an optimal {\em adaptive} strategy for ordering the tests used to evaluate the read-once formula. In an adaptive strategy,
the choice of the next test can depend on the outcomes of the previous tests, and thus the testing order corresponds to a decision tree, rather than to a permutation.  
It is still an open question whether the adaptive version of the problem is
 NP-hard, or whether it can be solved by a polynomial-time algorithm, even in the unit-cost case.   It is also unknown, even in the unit-cost case, whether it can be solved by a polynomial-time algorithm achieving a sublinear approximation factor.  In contrast, we establish here that there is a polynomial-time constant-factor approximation algorithm for the na-ROF evaluation problem in the unit-cost case.  However, as with the adaptive version of the problem, it remains open whether the na-ROF evaluation problem is, in fact, NP-hard.

An easy case of the na-ROF evaluation problem is where $\phi$ is the Boolean OR function, $\phi(x_1, \ldots, x_n) = x_1 \vee \ldots \vee x_n$.  In this case, the optimal solution is to
 perform the tests in decreasing order of the ratio $p_i/c_i$ until the value of $f$ can be determined (which occurs as soon as an $x_i=1$ is found, or after all $x_i$ have been found to equal 0).   
 This optimal solution has been rediscovered many times (cf.\ \cite{Unyulurt04}).  It is also optimal for the adaptive version of the problem.  
However, in contrast to the case of the OR function, an optimal linear order for evaluating a read-once formula will generally incur higher expected cost than an optimal adaptive strategy.  This is because, for example, learning that a variable $x_i=0$ when its parent node is labeled AND allows us to ``prune'' all other subtrees of that AND node, making it unnecessary to perform tests on any of the variables that were in the pruned subtrees.

\cite{BU00} and \cite{GU13} considered what might be called a ``partially-adaptive'' version of the na-ROF evaluation problem, where the strategy is specified by a permutation, but tests in the permutation are skipped if the results of previous tests have rendered them irrelevant.  They presented results on evaluating read-once formulas of small depth.
There does not appear to be a way to express this partially-adaptive version of the na-ROF evaluation problem in MSOP form.  

\cite{charikarPricedInfo}
studied the evaluation problem for read-once formulas in the worst-case online (non-stochastic) setting, where the goal is to minimize the so-called ``competitive ratio''.  
They gave an interesting exact algorithm for minimizing the competitive ratio, with pseudo-polynomial running time.  It is not applicable to our problem, where the goal is to minimize expected cost.

\subsection{Preliminaries}

Recall the inputs to the  na-ROF evaluation problem: (1) a read-once formula $\phi(x_1,\ldots,x_n)$, (2) for each $x_i$, the value $p_i := P[x_i=1]$ where $0 < p_i < 1$, and (3) for each $x_i$, the associated integer test cost $c_i$, which is greater than $0$. 
We may assume without loss of generality that each AND and OR gate of $\phi$ has exactly two inputs
(since, e.g., $x_1 \wedge x_2 \wedge x_3 = (x_1 \wedge x_2) \wedge x_3$).  We consider each input $x_i$ in $\phi$
to also be a gate (an {\em input gate}) of $\phi$.   The set of tests is $V = \{1, \ldots, n\}$. 

We use partial assignments to represent the outcomes of a subset of the tests.  In a partial assignment $b \in \{0,1,*\}^n$, $b_i = *$ means that test $i$ has not been performed and the value of $x_i$ is unknown, otherwise $b_i$ is the outcome of test $i$.  
For a (full) assignment $a \in \{0,1\}^n$ and $S$ a subset of $V = \{1, \ldots, n\}$, $a|_S$ is the partial assignment $b \in \{0,1,*\}^n$ where $b_i = a_i$ for $i \in S$, and $b_i = *$ otherwise.
Given partial assignment $b \in \{0,1,*\}^n$, an {\em extension} of $b$ is a (full) assignment $a \in \{0,1\}^n$ where $a_i = b_i$ for all $i$ such that $b_i \neq *$.  If for all extensions $a$ of $b$, $\phi(a)$ has the same value $\ell$, the value of $\phi$ {\em is determined by} $b$ and we write $\phi(b) = \ell$.  Otherwise, we write $\phi(b) = *$.
Intuitively, for $S \subset V$ and $a \in \{0,1\}^n$, 
$\phi(a|_S) = *$ means that the outcomes of the tests in $S$, as specified by $a$, are not sufficient to determine the value of $\phi$.

Let $A_1, \ldots, A_n$ be independent Bernoulli random variables, where
$P[A_i = 1]=p_i$.
Thus $A = [A_1, \ldots, A_n]$ is a random variable which takes on a value $a \in \{0,1\}^n$, corresponding to the outcome of the $n$ tests.  

In the MSOP formulation of the na-ROF evaluation problem, given above, we defined
$g(S)$ to be equal to the probability that the value of $\phi$ can be determined from the outcomes of the tests in $S$. Thus, $g(S)=P[\phi(A|_S) \neq *]$.

We will obtain an 8-approximate solution to the problem by constructing a $2$-greedy chain.  In particular, we will give a (pseudo) polynomial time 2-approximation algorithm solving the associated density problem.

\subsection{Background for the Dynamic Programming Algorithm}

The dynamic programming algorithm relies on the following definitions and observations.
For $S \subset T \subseteq V$, let
\begin{equation}
\rho_S(T) = \frac{P[\phi(A|_T) \neq *] - P[\phi(A|_S) \neq *]}{\sum_{i \in T \backslash S} c_i}.
\end{equation}

For $S \subset V$ and $\alpha > 0$, call $R \subseteq V \backslash S$ an $\alpha$-approximate max-density {\em supplement} for $S$ if
$$\rho_S(S \cup R) \geq \frac{1}{\alpha}\max_{\{T \subseteq V: S \subset T\}} \rho_S(T).$$

Our dynamic programming algorithm  computes a 2-approximate max-density supplement for an input subset $S$.  
It does so in a bottom-up fashion, calculating values at each of the gates $G$ of $\phi$.  
For gate $G$ of $\phi$, define $tests(G)$ to be the set of $i \in V$  such that $x_i$ is a descendant of $G$ in the tree $\phi$.
We consider a gate to be its own descendant, so if $G$ is an input gate $x_i$, then $i \in tests(G)$.


Each gate $G$ of $\phi$ is the root of a subtree of $\phi$.  Define $\phi_G$ to be the subformula corresponding to the subtree of $\phi$ that is rooted at $G$.  
Thus $\phi_G$ is a read-once formula over the variable  set $\{x_i  \mid i \in tests(G)\}$.  We treat $\phi_G$ as computing a function over $\{0,1\}^n$, whose output depends only on the values of the variables in $\{ x_i \mid i \in tests(G)\}$.  For $b \in \{0,1,*\}^n$, we refer to $\phi_G(b)$ as the output of gate $G$ on partial assignment $b$, which may be either 0, 1, or $*$. 

We note that given a subset $S \subseteq V$, and $\ell \in \{0,1\}$, the value of $P[\phi_G(A|_S)=\ell]$ for each gate $G$ of $\phi$ can be computed in time linear in $n$ by processing the gates of $\phi$ in bottom-up order.  Consider the case where $\ell=1$ and let $p_G = P[\phi_G(A|_S)=1]$.
If $G$ is an input gate $x_i$, then the value of $p_G$ depends on whether $x_i \in S$: if $x_i \in S$ then the value assigned to $x_i$ by $A|_S$ is $A_i$, so $p_G = P[A_i=1] = p_i$, but if $x_i \not\in S$, then the value assigned to $x_i$ by $A|_S$ is $*$, so $p_G = 0$.
If $G$ is an AND gate with children $G'$ and $G''$,
then because $\phi$ is read-once and the $A_i$ are independent, $p_G = p_{G'} \cdot p_{G''}$.  If $G$ is an OR gate, then $p_G = p_{G'} + p_{G''} - p_{G'} \cdot p_{G''}$.

Dually, consider the case where $\ell=0$ and let $q_G = P[\phi_G(A_S)=0]$.  
If $G$ is an input gate $x_i$, then $q_G = 1-p_i$ if $x_i \in S$, otherwise $q_G=0$.
If $G$ is an OR gate with children $G'$ and $G''$,
then $q_G = q_{G'}q_{G''}$.  If $G$ is an AND gate, then $q_G = q_{G'} + q_{G''} - q_{G'}q_{G''}$.

\subsection{The na-ROF Evaluation Algorithm}

Our algorithm for na-ROF evaluation relies on the dynamic programming algorithm presented in the proof of the following lemma.

\begin{lemma}
\label{lem:rof}
Given $S \subset V$,
a 2-approximate max-density supplement $R$ for $S$ can be computed in time polynomial in $n$ and  $\sum_{i=1}^n c_i$.
\end{lemma}

\textit{Proof.}
We prove the lemma for the case of unit costs, where all the $c_i$'s are equal to 1.  We then explain how to extend the proof to handle arbitrary costs.   

Assume the $c_i$'s  are all equal to 1. 
We describe an algorithm that we call {\tt FindSupp} that finds a max-density subset $R$ for a given input subset $S$.

Fix $S$. 
For $R \subseteq V \backslash S$, let
$\sigma(R) = \rho_S(S \cup R)$.  Since we assumed the $c_i$'s are equal to 1,
$$\sigma(R) = \frac{P[\phi(A|_{S \cup R}) \neq *] - P[\phi(A|_S) \neq *]}{|R|}.$$
Clearly, $P[\phi(A|_S) \neq *] = P[\phi(A|_S) = 1] + P[\phi(A|_S) = 0]$ and similarly for $A|_{S \cup R}$.  

For $\ell \in \{0,1\}$, define $$\sigma_{\ell}(R) = \frac{P[\phi(A|_{S \cup R}) = \ell] - P[\phi(A|_S) = \ell]}{|R|}.$$  Thus
\begin{equation}
    \label{eqsplit}
    \sigma(R) = \sigma_1(R) + \sigma_0(R)
\end{equation}

The idea behind {\tt FindSupp} is to compute two subsets $R^1$ and $R^0$, maximizing $\sigma_1$ and $\sigma_0$ respectively.  By (\ref{eqsplit}), the $R^{\ell}$ with the larger value of $\sigma_{\ell}(R^{\ell})$ is a 2-approximate max-density supplement for $S$.

For gate $G$ of $\phi$,  let $T(G) = tests(G) \backslash S$.  
For
$t  \in \{0, \ldots, |T(G)|\}$, and $\ell \in \{0,1\}$,
let $R_{G,t,\ell}$ be a subset $R$ that maximizes the value of $P[\phi_G(A|_{S \cup R}) = \ell]$ subject to the constraints that $R \subseteq T(G)$ and
$|R| = t$.   

Let $p_{G,t,\ell}$ be the value of
$P[\phi_G(A|_{S \cup R}) = \ell]$ for $R = R_{G,t,\ell}$.  Let $\tilde{G}$ denote the root gate of $\phi$.  Thus for $t \in \{1, \ldots, |V\backslash S|\}$ and $\ell \in \{0,1\}$,
setting $R=R_{\tilde{G},t,\ell}$ maximizes the value of $\sigma_{\ell}(R)$, over all $R \subseteq V \backslash S$ of size $t$.

{\bf \noindent Algorithm {\tt FindSupp}:}

{\tt FindSupp} first runs a procedure {\tt ComputeRp} that computes $R_{\tilde{G},t,\ell}$ and $p_{\tilde{G},t,\ell}$ for all $t \in \{1, \ldots, |V \backslash S|\}$ and $\ell \in \{0,1\}$. 
We describe the details of {\tt ComputeRp} below.
After running {\tt ComputeRp}, {\tt FindSupp} uses it to obtain the two subsets, $R^1$ and $R^0$, maximizing $\sigma_1$ and $\sigma_0$ respectively.
It does this as follows.
First, using the linear-time procedure described above, it computes the value of $P[\phi(A|_S) = \ell]$, for $\ell \in \{0,1\}$.  

Then, for each $t \in \{1, \ldots, |V \backslash S|\}$, for $\ell \in \{0,1\}$, {\tt FindSupp} computes the value of  $$\sigma_{\ell}(R_{\tilde{G},t,\ell}) = \frac{p_{\tilde{G},t,\ell} - P[\phi(A|_S) = \ell]}{t}$$

For each $\ell \in \{0,1\}$, the algorithm then finds the value of $t$ which yielded
the highest value for $\sigma_{\ell}(R_{\tilde{G},t,\ell})$.  Let $t^{\ell}$ denote that value.
Let $R^{\ell}$ be the value of  $R_{\tilde{G},t, \ell}$ for $t = t^{\ell}$.

Because $R_{\tilde{G},t,\ell}$ maximizes $\sigma_{\ell}$ among candidate subsets of size $t$,
setting $R=R^{\ell}$ maximizes $\sigma_{\ell}(R)$ among candidate subsets of all possible sizes.  
The algorithm returns $R^0$ if
$\sigma_{0}(R^{0}) > \sigma_{1}(R^1)$, and returns $R^1$ otherwise.

{\noindent \bf Procedure {\tt ComputeRp}:}

For all gates $G$ of $\phi$, 
{\tt ComputeRp} computes the values of $p_{G,t,\ell}$ and $R_{G,t,\ell}$ for all $t \in \{0, \ldots, |T(G)|\}$, and $\ell \in \{0,1\}$.  It processes the gates $G$ of $\phi$ in bottom-up order, from the leaves to the root.

We begin by describing how {\tt ComputeRp} computes $p_{G,t,\ell}$ and $R_{G,t,\ell}$ when $G$ is an AND gate, $t \in \{0, \ldots, |T(G)|\}$, and $\ell=1$.
Suppose that $G'$ and $G''$ are the children of AND gate $G$, and that
$R_{G',t',1}$, $p_{G',t',1}$, $R_{G'',t'',1}$, and $p_{G'',t'',1}$ have already been computed, for all $t' \in \{0, \ldots, |T(G')| \}$ and $t'' \in \{0, \ldots, |T(G'')| \}$.  
{\tt ComputeRp} first computes the product
 $p_{G',j,1} \cdot p_{G'',t-j,1}$ for all $j$ such that $j \in \{0, \ldots, |T(G')|\}$ and $t-j  \in \{0, \ldots, |T(G'')|\}$.
 It then sets $j^*$ to be the value of $j$ maximizing that product, 
 and sets
 $R_{G,t,1} = R_{G',j^*,1} \cup R_{G'',t-j^*,1}$ and $p_{G,t,1} = p_{G',j^*,1} \cdot p_{G'',t-j^*,1}$.
 
 The correctness of these settings follows from the fact that that $R_{G,t,1}$ must
consist of a subset $R'$ of $T(G')$ of some size $j^*$, and a subset $R''$
of $T(G'')$ of size $t-j^*$.  
$R_{G,t,1}$ maximizes the probability that $G$ outputs 1 (among subsets of $T(G)$ of size $t$, when added to $S$).  Since $\phi$ is a read-once formula, $tests(G')$ and $tests(G'')$ are disjoint.  
$R'$ and $R''$ are thus sets that maximize the probability that $G'$ and $G''$ output 1 (when added to $S$, among subsets of  $T(G')$ and $T(G'')$ of sizes $j^*$ and $t-j^*$ respectively). 
Thus, given $j^*$, $R_{G,t,1}$ can be set to $R' \cup R''$ where
$R' = R_{G',j^*,1}$, $R''= R_{G'',t-j^*,1}$,
and $p_{G,t,1}$ can be set to $p_{G',j^*,1} \cdot p_{G'',t-j^*,1}$.
Since {\tt ComputeRp} is not given the value of $j^*$, it must try all possible $j$.  
 
Similarly, suppose $G$ is an AND gate, $t \in \{0, \ldots, |T(G)|\}$, and $\ell=0$. 
In this case,  {\tt ComputeRp}
computes
$p_{G',j,0} + p_{G'',t-j,0} - p_{G',j,0} \cdot p_{G''.t-j,0}$
for all possible $j$, and then sets $j^*$ to be the value that maximized
the expression.  It then sets
$R_{G,t,0} = R_{G',j^*,0} \cup R_{G'',t-j^*,0}$ and $p_{G,t,0} = p_{G',j^*,0} + p_{G'',t-j^*,0} - p_{G',j^*,0} \cdot p_{G'',t-j^*,0}$.
The correctness in this case follows from the fact that the output of $G$ is 0 if either of its child gates outputs 0.  Thus  to maximize the probability that AND gate $G$ outputs 0, one needs to maximize the probability that each of its child gates outputs 0.

The case where $G$ is an OR gate is dual and we omit the details.  

The remaining case is where $G$ is an input gate  $x_i$, $t \in \{0, \ldots, T(G)\}$, and $\ell \in \{0,1\}$.  Note that since $G$ is an input gate,
if $i \in S$, then $|T(G)| =0$.  If $i \not\in S$, then $|T(G)| = 1$.  If $t=0$, then for $\ell \in \{0,1\}$, {\tt ComputeRp} 
sets $R_{G,t,\ell} = \emptyset$.  Then,
if $i \in S$ it sets 
$p_{G,t,1} = p_i$ and
$p_{G,t,0} = 1-p_i$.  If $i \not\in S$
it sets both $p_{G,t,0}=0$ and $p_{G,t,1}  = 0$. 
If $t=1$ (and therefore $i \not\in S$), it sets 
$R_{G,t,\ell} = x_i$,  $p_{G,t,1} = p_i$ and
$p_{G,t,0} = 1-p_i$.  The correctness of these settings is straightforward.

To compute the running time of {\tt ComputeRp}, note that because $\phi$ is a read-once formula with $n$ input variables, it has $\mathcal{O}(n)$ gates.
For each gate $G$, $|T(G)|$ is $\mathcal{O}(n)$, and thus there are $\mathcal{O}(n^2)$ values 
computed by {\tt ComputeRp}.
For each $G,t,\ell$ where $G$ is an AND or OR gate, {\tt ComputeRp} spends time linear
in $|T(G)|$ to find $j^*$, which
yields the values for $p_{G,t,\ell}$ and $R_{G,t,\ell}$.  Thus the running time of
{\tt ComputeRp} is $\mathcal{O}(n^3)$.

%

%
%

 
{\bf \noindent Generalization to arbitrary costs:}

The algorithm {\tt FindSupp} can easily be modified to handle arbitrary non-negative integer costs.
The main difference is that
$t$ is used to represent the total cost of a set $R$ of tests,
rather than just the size of the set.

Consider a gate $G$ of $\phi$.
If there is at least one subset $R \subseteq T(G)$ such that $t = \sum_{i \in R} c_i$, call $t$ a feasible value for $G$. 

For $t$ a feasible value for $G$, let
$R_{G,t,\ell}$ be the subset maximizing 
  $\rho_R(S)$ (whose denominator is now $\sum_{i \in R} c_i$)
 subject to the constraints that $R \subseteq T(G)$ and $\sum_{i \in R} c_i = t$. 
 Because of our assumption that the costs $c_i$ are integers, the feasible values of $t$ are all in the set $\{1, \ldots, \sum_{i \in V} c_i\}$.
 
 {\tt ComputeRp} computes $R_{G,t,\ell}$ for all feasible values $t$ for $G$, as follows.
 Suppose $G$ is an AND or OR gate.   Let $G'$ and $G''$ be its children.
  {\tt ComputeRp} identifies the feasible values $t$ for $G$, which are the values $t = t' + t''$ such that $t'$ is feasible for $G'$ and $t''$ is feasible for $G''$.
 To compute $R_{G,t,\ell}$
 for a feasible value $t$ for $G$,  
 {\tt ComputeRp} tries all $j$
such that $j$ is feasible for $G'$ and $t-j$ is feasible for $G''$.  
The other modifications in the algorithm are straightforward.  

Because there are at most $\sum_{i \in V} c_i$ feasible values $t$ for  gates $G$,
the running time of {\tt FindSupp} is $\mathcal{O}(n(\sum_{i \in V} c_i)^2)$. 
$\Box$

The algorithm described in Lemma~\ref{lem:rof} can be used to
form a $2$-greedy chain $S_0 \subset S_1 \subset \ldots \subset S_k$, where each $S_{j+1}$ is generated from $S_j$ by running {\tt FindSupp} with $S = S_j$ to produce $R$, and then setting $S_{j+1} = S_j \cup R$.

The theorem now follows immediately from Lemma~\ref{lem:rof} and Theorem~\ref{thm:4-approxforward}.

\begin{theorem}
There is a pseudo-polynomial time 8-approximation algorithm solving the na-ROF evaluation problem. The algorithm runs in polynomial time in the unit cost case.
\end{theorem}

\section{Backward Greedy Algorithms} \label{sec:dual}



We call an $\mathcal{F}$-chain $\mathcal{S}=(S_j)_{j=0}^k$ a {\em backward $\alpha$-greedy chain} if
\[
\rho_{S_{j}}(S_{j-1}) \le \alpha  \min_{\{T \in \mathcal{F}: T \subseteq S_{j} \}} \rho_{S_j}(T),
\]
for all $j=1,\ldots, k$.

A backward $\alpha$-greedy chain can be seen to be equivalent to an $\alpha$-greedy chain for a dual version of the MSOP problem.  To describe this dual problem, we write the cost function $C(f,g)$ in another, equivalent way.  
Fixing $\mathcal{F} \subseteq 2^V$, we first define $\mathcal{F}^\#$ as the family of complements of sets in $\mathcal{F}$. 
That is, $\mathcal{F}^\#=\{S\subseteq V: V \setminus S \in \mathcal{F}\}$. 
Note that $\mathcal{F}$ is closed under intersection if and only if $\mathcal{F}^\#$ is closed under union. 
There is a one-to-one correspondence between $\mathcal{F}$-chains and $\mathcal{F}^\#$-chains, obtained by mapping an element $S$ of an $\mathcal{F}$-chain to $V \setminus S$ and reversing the order. We refer to the corresponding $\mathcal{F}^\#$-chain of an $\mathcal{F}$-chain $\mathcal{S}$ as its {\em dual chain}, which we denote by $\mathcal{S}^\#$. 
We also denote the {\em dual function} of $f$ by $f^\#:\mathcal{F}^\# \rightarrow \mathbb{R}$, given by $f^\#(S)=f(V)-f(V \setminus S)$, and similarly for $g$. Note that a set function is the dual of its dual, as is an $\mathcal{F}$-chain, and that $f^\#$ and $g^\#$ are non-decreasing. 
Also, $f$ is submodular if and only if $f^\#$ is supermodular.

Given an MSOP with inputs $f$, $g$ and $\mathcal{F}$, the dual problem is an MSOP with inputs $g^\#$, $f^\#$ and $\mathcal{F}^\#$. In other words, the dual problem is to minimize
\begin{equation*}
C_{g^\#,f^\#}(\mathcal{T}) = \sum_{j=1}^k g^\#(T_j)(f^\#(T_j)-f^\#(T_{j-1})), 
\end{equation*}
over all $\mathcal{F}^\#$-chains $\mathcal{T}=(T_j)_{j=0}^k$. Observe that an MSOP is the dual of its dual. 

It is now easy to see that an $\mathcal{F}$-chain $\mathcal{S}=(S_j)_{j=0}^n$ is a backward $\alpha$-greedy chain if and only if its dual chain is an $\alpha$-greedy $\mathcal{F}^\#$-chain for the dual problem.

The following is immediate and generalizes a similar observation from \cite{FLV19}. 
\begin{lemma} \label{lem:dual}
If $\mathcal{S}$ is an $\mathcal{F}$-chain, then $C_{f,g}(\mathcal{S})=C_{g^\#,f^\#}(\mathcal{S^\#})$ and $\mathcal{S}$ is an $\alpha$-approximation for an instance of MSOP if and only if $\mathcal{S}^\#$ is an $\alpha$-approximation for its dual. 
\end{lemma}
\textit{Proof.}
To prove the first statement, we point out that the area $A$ under the second histogram in the proof of Theorem~\ref{thm:4-approxforward}, given by the sum in~(\ref{eq:dual}), is equal to $C_{g^\#,f^\#}(\mathcal{S^\#})$. This area is also shown to be equal to $C_{f,g}(\mathcal{S})$ later in the same proof. The second statement in the lemma follows directly from the first. $\Box$

Applying Theorem~\ref{thm:4-approxforward} to the dual problem, we obtain the following theorem as a corollary. 

\begin{theorem} \label{thm:4-approxback}
Suppose $g^\#$ is subadditive and $\mathcal{F}$ is closed under intersection. Then for any $\alpha \ge 1$, a backward $\alpha$-greedy chain is a $4 \alpha$-approximation for an optimal chain for MSOP.
\end{theorem}

Note that if $g$ is supermodular, then $g^\#$ is submodular and hence subadditive, and therefore satisfies the hypothesis of Theorem~\ref{thm:4-approxback}. 

We may also apply Corollary~\ref{cor:4approx} to the dual problem to obtain an additional corollary.

\begin{corollary} \label{cor:4approx2}
Suppose $\mathcal{F}=2^V$. If $f$ is supermodular and $g$ is modular then a backward {$1 \text{-greedy}$} chain can be found in polynomial time and there exists a polynomial time $4$-approximation algorithm for MSOP.
\end{corollary}

\section{Future Work} \label{sec:conclusion}

We have created a general framework for min-sum ordering problems, and while Theorem~\ref{thm:4-approxforward} relies on very modest assumptions, it is only useful if the maximum density problem can be efficiently approximated. More work is needed in this area in order to further exploit our approximation result. 

Particular problems of interest include Generalized Min Sum Set Cover (GMSSC), introduced in \cite{azar2009multiple}. Unlike MSSC, where a hyperedge is ``covered'' the first time any of its vertices are chosen, in GMSSC, each hyperedge has its own ``covering requirement'', which specifies how many of its vertices must be chosen before it is ``covered''. The objective is to minimize the sum of covering times, as in MSSC. 
If the associated maximum density problem could be approximated within a factor of $\alpha$, this would yield a $4\alpha$-approximation algorithm for GMSSC.  
The best approximation factor obtained to date for GMSSC is 4.642, due to~\cite{bansal2021improved}.  Their algorithm solves an LP relaxation and then 
applies a kernel transformation and randomized rounding.  
By adding costs to vertices and weights to hyperedges, one could further generalize GMSSC, giving rise to a more general maximum density problem of interest.

Another problem that comes under our framework is the Unreliable Job Scheduling Problem (UJP), introduced in \cite{agnetis2009sequencing}. In the basic setting, a set of jobs with given rewards must be scheduled by a single machine to maximize the total expected reward. There is a probability of failure associated with each job when it is scheduled, and if failure occurs the machine cannot schedule any further jobs. The problem has a neat ``index'' solution. Natural generalizations of the problem would consider the possibility of AND- or OR-precedence constraints, and therefore their associated maximum density problems.

\section*{Acknowledgements} The authors would like to thank Andreas S. Schulz for helpful comments in improving the presentation of the paper. This material is based upon work supported by the National Science Foundation under Grant Numbers IIS-1909446 and IIS-1909335 and by the Alexander von Humboldt Foundation with funds from the German Federal Ministry of Education and Research (BMBF). 

\bibliographystyle{apalike}
\bibliography{references}

\begin{thebibliography}{}

\bibitem[Agnetis et~al., 2009]{agnetis2009sequencing}
Agnetis, A., Detti, P., Pranzo, M., and Sodhi, M.~S. (2009).
\newblock Sequencing unreliable jobs on parallel machines.
\newblock {\em Journal of Scheduling}, 12(1):45.

\bibitem[Alpern and Lidbetter, 2013]{AL13}
Alpern, S. and Lidbetter, T. (2013).
\newblock Mining coal or finding terrorists: The expanding search paradigm.
\newblock {\em Operations Research}, 61(2):265--279.

\bibitem[Amb{\"u}hl and Mastrolilli, 2009]{AM09}
Amb{\"u}hl, C. and Mastrolilli, M. (2009).
\newblock Single machine precedence constrained scheduling is a vertex cover
  problem.
\newblock {\em Algorithmica}, 53(4):488--503.

\bibitem[Averbakh and Pereira, 2012]{AP12}
Averbakh, I. and Pereira, J. (2012).
\newblock The flowtime network construction problem.
\newblock {\em IIE Transactions}, 44(8):681--694.

\bibitem[Azar et~al., 2009]{azar2009multiple}
Azar, Y., Gamzu, I., and Yin, X. (2009).
\newblock Multiple intents re-ranking.
\newblock In {\em Proceedings of the Forty-First Annual ACM Symposium on Theory
  of Computing}, pages 669--678.

\bibitem[Bansal et~al., 2021]{bansal2021improved}
Bansal, N., Batra, J., Farhadi, M., and Tetali, P. (2021).
\newblock Improved approximations for min sum vertex cover and generalized min
  sum set cover.
\newblock In {\em Proceedings of the 2021 ACM-SIAM Symposium on Discrete
  Algorithms (SODA)}, pages 998--1005. SIAM.

\bibitem[Bar-Noy et~al., 1998]{BNBHST98}
Bar-Noy, A., Bellare, M., Halld{\'o}rsson, M.~M., Shachnai, H., and Tamir, T.
  (1998).
\newblock On chromatic sums and distributed resource allocation.
\newblock {\em Information and Computation}, 140(2):183--202.

\bibitem[Bar-Noy and Kortsarz, 1998]{BNK98}
Bar-Noy, A. and Kortsarz, G. (1998).
\newblock Minimum color sum of bipartite graphs.
\newblock {\em Journal of Algorithms}, 28(2):339--365.

\bibitem[Boros and {\"U}nl{\"u}yurt, 2000]{BU00}
Boros, E. and {\"U}nl{\"u}yurt, T. (2000).
\newblock Sequential testing of series-parallel systems of small depth.
\newblock In Laguna, M. and Velarde, J. L.~G., editors, {\em Computing Tools
  for Modeling, Optimization and Simulation: Interfaces in Computer Science and
  Operations Research}, pages 39--73. Springer US, Boston, MA.

\bibitem[Charikar et~al., 2000]{charikarPricedInfo}
Charikar, M., Fagin, R., Guruswami, V., Kleinberg, J., Raghavan, P., and Sahai,
  A. (2000).
\newblock Query strategies for priced information (extended abstract).
\newblock In {\em Proceedings of the 32nd annual ACM Symposium on Theory of
  Computing}, STOC '00, pages 582--591, New York, NY, USA. ACM.

\bibitem[Chekuri and Motwani, 1999]{CM99}
Chekuri, C. and Motwani, R. (1999).
\newblock Precedence constrained scheduling to minimize sum of weighted
  completion times on a single machine.
\newblock {\em Disc. Appl. Math.}, 98(1):29--38.

\bibitem[Chudak and Hochbaum, 1999]{CH99}
Chudak, F.~A. and Hochbaum, D.~S. (1999).
\newblock A half-integral linear programming relaxation for scheduling
  precedence-constrained jobs on a single machine.
\newblock {\em Operations Research Letters}, 25(5):199--204.

\bibitem[Correa and Schulz, 2005]{CS05}
Correa, J.~R. and Schulz, A.~S. (2005).
\newblock Single-machine scheduling with precedence constraints.
\newblock {\em Mathematics of Operations Research}, 30(4):1005--1021.

\bibitem[Deshpande et~al., 2016]{DHK16}
Deshpande, A., Hellerstein, L., and Kletenik, D. (2016).
\newblock Approximation algorithms for stochastic submodular set cover with
  applications to boolean function evaluation and min-knapsack.
\newblock {\em {ACM} Trans. Algorithms}, 12(3):42:1--42:28.

\bibitem[Erlebach et~al., 2003]{EKM03}
Erlebach, T., K{\"a}{\"a}b, V., and M{\"o}hring, R.~H. (2003).
\newblock Scheduling {AND/OR}-networks on identical parallel machines.
\newblock In {\em International Workshop on Approximation and Online
  Algorithms}, volume 2909 of {\em LNCS}, pages 123--136. Springer.

\bibitem[Feige et~al., 2002]{FLT02}
Feige, U., Lov{\'a}sz, L., and Tetali, P. (2002).
\newblock Approximating min-sum set cover.
\newblock In {\em International Workshop on Approximation Algorithms for
  Combinatorial Optimization}, volume 2462 of {\em LNCS}, pages 94--107.
  Springer.

\bibitem[Feige et~al., 2004]{FLT04}
Feige, U., Lov{\'a}sz, L., and Tetali, P. (2004).
\newblock Approximating min sum set cover.
\newblock {\em Algorithmica}, 40(4):219--234.

\bibitem[Fokkink et~al., 2017]{FKR17}
Fokkink, R., Kikuta, K., and Ramsey, D. (2017).
\newblock The search value of a set.
\newblock {\em Annals of Oper. Res.}, 256(1):63--73.

\bibitem[Fokkink et~al., 2019]{FLV19}
Fokkink, R., Lidbetter, T., and V\'{e}gh, L. (2019).
\newblock On submodular search and machine scheduling.
\newblock {\em Mathematics of Operations Research}, 44(4):1431--1449.

\bibitem[Furnas and Zacks, 1994]{furnas1994multitrees}
Furnas, G.~W. and Zacks, J. (1994).
\newblock Multitrees: enriching and reusing hierarchical structure.
\newblock In {\em Proceedings of the SIGCHI Conference on Human Factors in
  Computing Systems}, pages 330--336.

\bibitem[Gillies and Liu, 1995]{GL95}
Gillies, D.~W. and Liu, J. W.-S. (1995).
\newblock Scheduling tasks with and/or precedence constraints.
\newblock {\em SIAM Journal on Computing}, 24(4):797--810.

\bibitem[Gkenosis et~al., 2018]{GGHK18}
Gkenosis, D., Grammel, N., Hellerstein, L., and Kletenik, D. (2018).
\newblock The stochastic score classification problem.
\newblock In {\em Proceedings of the 26th Annual European Symposium on
  Algorithms, {ESA}}, pages 36:1--36:14.

\bibitem[Greiner et~al., 2006]{GHJM06}
Greiner, R., Hayward, R., Jankowska, M., and Molloy, M. (2006).
\newblock Finding optimal satisficing strategies for and-or trees.
\newblock {\em Artificial Intelligence}, 170(1):19--58.

\bibitem[Happach and Schulz, 2020a]{HS20arxiv}
Happach, F. and Schulz, A.~S. (2020a).
\newblock Approximation algorithms and {LP} relaxations for scheduling problems
  related to min-sum set cover.
\newblock {\em arXiv:2001.07011}.

\bibitem[Happach and Schulz, 2020b]{HS20waoa}
Happach, F. and Schulz, A.~S. (2020b).
\newblock Precedence-constrained scheduling and min-sum set cover.
\newblock In {\em Approximation and Online Algorithms}, volume 11926 of {\em
  LNCS}, pages 170--187. Springer.

\bibitem[Hermans et~al., 2021]{hermans2021exact}
Hermans, B., Leus, R., and Matuschke, J. (2021).
\newblock Exact and approximation algorithms for the expanding search problem.
\newblock {\em INFORMS Journal on Computing}.

\bibitem[Horn, 1972]{horn1972single}
Horn, W. (1972).
\newblock Single-machine job sequencing with treelike precedence ordering and
  linear delay penalties.
\newblock {\em SIAM Journal on Applied Mathematics}, 23(2):189--202.

\bibitem[I{\c s}{\i}k and {\"U}nl{\"u}yurt, 2013]{GU13}
I{\c s}{\i}k, G. and {\"U}nl{\"u}yurt, T. (2013).
\newblock Sequential testing of 3-level deep series-parallel systems.
\newblock In {\em IEEE International Conference on Industrial Engineering and
  Engineering Management}.

\bibitem[Iwata et~al., 2012]{ITT12}
Iwata, S., Tetali, P., and Tripathi, P. (2012).
\newblock Approximating minimum linear ordering problems.
\newblock In {\em Approximation, Randomization, and Combinatorial Optimization.
  Algorithms and Techniques}, pages 206--217. Springer.

\bibitem[Johannes, 2005]{J05}
Johannes, B. (2005).
\newblock On the complexity of scheduling unit-time jobs with {OR}-precedence
  constraints.
\newblock {\em Operations Research Letters}, 33(6):587--596.

\bibitem[Kubicka and Schwenk, 1989]{KS89}
Kubicka, E. and Schwenk, A.~J. (1989).
\newblock An introduction to chromatic sums.
\newblock In {\em Proceedings of the 17th conference on ACM Annual Computer
  Science Conference}, pages 39--45. ACM.

\bibitem[Lawler, 1978]{lawler1978sequencing}
Lawler, E.~L. (1978).
\newblock Sequencing jobs to minimize total weighted completion time subject to
  precedence constraints.
\newblock In {\em Annals of Discrete Mathematics}, volume~2, pages 75--90.
  Elsevier.

\bibitem[Margot et~al., 2003]{MQW03}
Margot, F., Queyranne, M., and Wang, Y. (2003).
\newblock Decompositions, network flows and a precedence constrained single
  machine scheduling problem.
\newblock {\em Oper. Res.}, 51(6):981--992.

\bibitem[McClintock et~al., 2017]{McCMW17}
McClintock, J., Mestre, J., and Wirth, A. (2017).
\newblock Precedence-constrained min sum set cover.
\newblock In {\em 28th International Symposium on Algorithms and Computation},
  volume~92 of {\em Leibniz International Proceedings in Informatics (LIPIcs)},
  pages 55:1--55:12.

\bibitem[Munagala et~al., 2005]{MBMW05}
Munagala, K., Babu, S., Motwani, R., and Widom, J. (2005).
\newblock The pipelined set cover problem.
\newblock In {\em International Conference on Database Theory}, volume 3363 of
  {\em LNCS}, pages 83--98. Springer.

\bibitem[Pisaruk, 1992]{Pisaruk}
Pisaruk, N. (1992).
\newblock The boundaries of submodular functions.
\newblock {\em Comp. Math. Math. Phys.}, 32(12):1769--1783.

\bibitem[Schulz, 1996]{Schulz}
Schulz, A.~S. (1996).
\newblock Scheduling to minimize total weighted completion time: Performance
  guarantees of {LP}-based heuristics and lower bounds.
\newblock In {\em Proceedings of the 5th International Conference on Integer
  Programming and Combinatorial Optimization}, volume 1084 of {\em LNCS}, pages
  301--315. Springer.

\bibitem[Sidney, 1975]{Sidney}
Sidney, J. (1975).
\newblock Decomposition algorithms for single-machine sequencing with
  precedence relations and deferral costs.
\newblock {\em Oper. Res.}, 23(2):283--298.

\bibitem[Streeter and Golovin, 2008]{DBLP:conf/nips/StreeterG08}
Streeter, M.~J. and Golovin, D. (2008).
\newblock An online algorithm for maximizing submodular functions.
\newblock In Koller, D., Schuurmans, D., Bengio, Y., and Bottou, L., editors,
  {\em Advances in Neural Information Processing Systems 21, Proceedings of the
  Twenty-Second Annual Conference on Neural Information Processing Systems,
  Vancouver, British Columbia, Canada, December 8-11, 2008}, pages 1577--1584.
  Curran Associates, Inc.

\bibitem[{\"U}nl{\"u}yurt, 2004]{Unyulurt04}
{\"U}nl{\"u}yurt, T. (2004).
\newblock Sequential testing of complex systems: a review.
\newblock {\em Discrete Applied Mathematics}, 142(1-3):189--205.

\end{thebibliography}

\end{document}